\documentclass[doublecolumn,11pt]{IEEEtran}
\usepackage{cite}
\usepackage{amsmath,amssymb,amsfonts}
\usepackage{graphicx,color}
\usepackage{textcomp}
\usepackage{hyperref}
\usepackage{enumitem}
\usepackage{multirow}
\usepackage{titlesec}
\usepackage{xcolor}
\usepackage{subcaption} 
\usepackage{placeins}
\usepackage{booktabs}
\usepackage{tabularx}
\usepackage{dblfloatfix} 

\usepackage[linesnumbered,ruled,vlined]{algorithm2e}

\def\BibTeX{{\rm B\kern-.05em{\sc i\kern-.025em b}\kern-.08em
    T\kern-.1667em\lower.7ex\hbox{E}\kern-.125emX}}
\begin{document}

\title{
\textbf{Benchmarking Deep Neural Networks for Modern Recommendation System} }

\author{
\textbf{Abderaouf Bahi\textsuperscript{a*}},
\textbf{Inoussa Mouiche\textsuperscript{b}} and
\textbf{Ibtissem Gasmi\textsuperscript{a}} \\ [1em]

\textsuperscript{a}\small
Computer Science and Applied Mathematics Laboratory (LIMA)\\ Faculty of Science and Technology, Chadli Bendjedid University, P.O. Box 73, El Tarf 36000, Algeria \\[0.6em]
\textsuperscript{b}\small
School of Computer Science, University of Windsor, ON, Canada\\[0.6em]

\textit{*Corresponding author:} Abderaouf Bahi (\textbf{a.bahi@univ-eltarf.dz})
}

\maketitle

\begin{abstract}
This paper presents a requirement-oriented benchmark of seven deep neural architectures, CNN, RNN, GNN, Autoencoder, Transformer, Neural Collaborative Filtering, and Siamese Networks, across three real-world datasets: Retail E-commerce, Amazon Products, and Netflix Prize. To ensure a fair and comprehensive comparison aligned with the evolving demands of modern recommendation systems, we adopt a Requirement-Oriented Benchmarking (ROB) framework that structures evaluation around predictive accuracy, recommendation diversity, relational awareness, temporal dynamics, and computational efficiency. Under a unified evaluation protocol, models are assessed using standard accuracy-oriented metrics alongside diversity and efficiency indicators. Experimental results show that different architectures exhibit complementary strengths across requirements, motivating the use of hybrid and ensemble designs. The findings provide practical guidance for selecting and combining neural architectures to better satisfy multi- objective recommendation system requirements.
\end{abstract}



\def\BibTeX{{\rm B\kern-.05em{\sc i\kern-.025em b}\kern-.08em
    T\kern-.1667em\lower.7ex\hbox{E}\kern-.125emX}}

\section{Introduction}
Technological advancements and evolving consumer behavior have driven an unprecedented expansion of the digital marketplace in recent years. In the first quarter of 2023 alone, online transactions increased by more than 8\% compared to the previous year, reaching over 540 million transactions and generating revenues exceeding 41 billion euros \cite{1,2,3}. This rapid growth underscores not only the scale of modern e-commerce platforms but also raises a critical question: how can digital systems effectively sustain user engagement and conversion in increasingly competitive and data-intensive environments?

Recommendation systems play a central role in addressing this challenge. By leveraging large volumes of user interaction data—such as preferences, purchase histories, and behavioral patterns. These systems aim to deliver personalized content that enhances user experience and drives sales \cite{4,5}. Beyond predictive accuracy, modern recommendation systems are increasingly expected to satisfy additional requirements, including recommendation diversity, relational awareness, temporal adaptability, and scalability. In particular, diversity has emerged as a key factor in mitigating informational lock-in, where users are repeatedly exposed to similar or overly popular items, thereby limiting discovery and long-term engagement  \cite{7,8,10}. Encouraging exploration through diverse recommendations has been shown to improve user satisfaction and retention \cite{13,14,15}.

Despite these advances, achieving an effective balance between accuracy and diversity remains a significant challenge \cite{16,18}. Moreover, different neural network architectures exhibit varying strengths in addressing these requirements, depending on how they model relationships, sequential behavior, or latent representations. While prior studies have explored individual neural architectures for recommendation tasks, evaluation practices often remain fragmented, focusing on isolated performance metrics without explicitly accounting for the multi-objective nature of modern recommendation systems. To address this limitation, we adopt a Requirement-Oriented Benchmarking (ROB) perspective, which frames recommendation evaluation around a set of core system requirements, including predictive accuracy, diversity, relational modeling capability, temporal adaptability, and computational efficiency. Rather than proposing new models or metrics, ROB provides a structured lens for systematically comparing existing architectures under a unified and application-relevant evaluation setting.

Under ROB, this study presents a comprehensive benchmark of seven neural network architectures, Convolutional Neural Networks (CNNs), Recurrent Neural Networks (RNNs), Graph Neural Networks (GNNs), Autoencoders, Transformers, Neural Collaborative Filtering (NCF), and Siamese Networks, for item–item recommendation tasks. Using three real-world datasets from retail e-commerce, online product platforms, and media consumption, the models are evaluated under a unified experimental protocol with respect to both predictive accuracy and recommendation diversity. The objective is to identify which architectures are best suited to specific system requirements and to provide practical guidance for designing recommendation systems that align with the evolving demands of modern digital platforms.

The remainder of the paper is organized as follows. Section 2 reviews related work on neural network–based recommendation systems. Section 3 describes the experimental methodology and evaluation protocol. Section 4 presents and discusses the comparative results. Section 5 concludes the paper and outlines future research directions.

\section{Related Works}

\subsection{Overview of Recommendation Systems}
Recommendation systems are widely used to improve user experience and engagement in e- commerce and digital content platforms, and they typically fall into collaborative filtering, content- based, and hybrid categories \cite{18,19,20,21}. Collaborative filtering leverages historical user–item interactions, user-based methods exploit similarity among users, and content-based approaches recommend items based on item attributes or user profiles \cite{22,23}. While these paradigms remain foundational, modern recommendation systems increasingly incorporate deep learning to address complex requirements such as relational modeling, temporal preference dynamics, and scalability.

Recent studies reflect this shift toward more sophisticated architectures and deployment settings. For example, Enqi Yu et al. \cite{25} proposed ClusterFedMet, a federated recommendation approach that integrates user clustering and meta-learning to improve personalization under Non-IID data distributions, while also reducing communication overhead. However, privacy and security concerns remain open challenges in federated environments. Similarly, Huiying Shi et al. \cite{26} combined Graph Neural Networks (GNNs) with collaborative filtering to capture richer interaction structures, though generalization across datasets with differing interaction attributes remains difficult. Other empirical work has highlighted the value of content-based filtering for long-tail and niche recommendation scenarios. Mohsen Jozani et al. \cite{27} showed that content-based filtering can promote less popular items and help distribute demand beyond highly frequent products, which implicitly relates to diversity and coverage objectives. Hybrid strategies have also been explored to improve relevance, such as combining sentiment signals with matrix factorization; X.J. Li et al. \cite{28} integrated LSTM-based sentiment analysis with matrix decomposition to enhance accuracy, but noted limitations in scalability and computational efficiency.

Overall, prior work demonstrates that recommendation systems must balance multiple objectives, including accuracy, diversity, temporal adaptability, and computational feasibility, yet evaluation practices are often fragmented and centered on limited metrics or single datasets. This motivates the need for systematic, requirement-oriented benchmarking under consistent experimental conditions, which is the focus of our study.

\subsection{Neural Network Architectures in Recommender Systems}

The adoption of neural network architectures has significantly advanced recommendation systems by enabling the modeling of nonlinear and complex user–item interaction patterns \cite{29,30,31}. Deep learning–based recommenders can process large-scale interaction data and capture latent representations that improve recommendation relevance beyond traditional methods \cite{32}. Consequently, a wide range of architectures, including convolutional, recurrent, graph-based, autoencoder-based, and attention-based models, have been explored for recommendation tasks.

Recent research has also investigated architectural efficiency and scalability as important considerations for deploying neural models in real-world systems. For example, studies on optimized neural architectures and model compression techniques have demonstrated that performance gains often come at the cost of increased computational complexity and careful hyperparameter tuning \cite{33,34,35}. While such works are not specific to recommendation tasks, they highlight broader challenges related to computational efficiency and scalability that are equally relevant for modern recommender systems. However, these studies typically focus on architectural optimization rather than comparative evaluation of recommendation performance across multiple system objectives.

\subsection{Accuracy-Oriented Evaluation in Recommendation Systems}
Predictive accuracy remains a primary evaluation criterion in recommendation system research, as it directly reflects the relevance of recommended items to user preferences. Numerous studies focus on optimizing accuracy metrics such as precision, recall, and F1-score across different datasets and application domains.
Lei Hou and Yichen Huang \cite{36} examined the impact of recommendation list length on system performance using datasets including Steam, MovieLens, and Amazon, revealing an inherent trade-off between accuracy and diversity as the recommendation list expands. Similarly, Muhammad Umar et al. \cite{37} analyzed recommendation accuracy in financial contexts, demonstrating that accuracy can vary substantially depending on task formulation and evaluation criteria. While these studies provide valuable insights into accuracy optimization, they often evaluate performance under isolated conditions and do not systematically account for other system requirements such as diversity, temporal dynamics, or computational efficiency.

\subsection{Diversity-Oriented Evaluation in Recommendation Systems}
Recommendation diversity is a critical objective in modern recommendation systems, as it promotes exposure to a broader range of items, reduces over-specialization, and mitigates echo chamber effects \cite{38,39,40}. By encouraging users to explore unfamiliar or less popular items, diversity can improve long-term engagement and user satisfaction \cite{41,42}. As a result, diversity has increasingly been studied as a complementary objective to predictive accuracy.

Several approaches have been proposed to explicitly enhance diversity in recommendation outputs. Dunlu Peng and Yi Zhou \cite{43} introduced LAP-SR, a post-processing framework that improves long-tail exposure in session-based recommendation through personalized diversity control. Zihao Li et al. \cite{44} proposed Teddy, a sequential recommendation model that disentangles user interest trends and diversity using temporal convolutional networks and multi-layer perceptrons, although its dual-pathway architecture introduces additional complexity that may limit real-time applicability. Huaizhen Kou et al. \cite{45} developed DI-RAR, a diversity-driven recommendation approach for APIs based on mashup graphs, but noted increased computational overhead as a practical limitation.

Beyond traditional recommendation scenarios, diversity has also been examined in specialized domains. Sofia Morgado Pereira et al. \cite{46} analyzed diversity in earthquake preparedness recommendations across European regions, revealing inconsistencies in information delivery, while Alvise De Biasio et al. \cite{47} proposed diversity-aware strategies to reduce harmful content exposure for sensitive users. Although these studies target different application contexts, they collectively highlight the growing importance of diversity as a system-level requirement.

Despite these advances, existing work often evaluates diversity in isolation or treats it as a secondary objective, limiting the ability to systematically analyze trade-offs between diversity, accuracy, and other system requirements under unified experimental settings.

\subsection{Positioning Our Work in the State of the Art}

Recent research demonstrates rapid progress in neural network–based recommendation systems, including graph-based approaches \cite{26}, sequence-aware models, and neural architecture optimization techniques \cite{34}. While these studies advance specific aspects of recommendation performance, they typically evaluate individual architectures under isolated objectives, limited datasets, or non-unified experimental conditions. In particular, relatively few works explicitly
examine the joint behavior of predictive accuracy and recommendation diversity in item–item recommendation settings across heterogeneous datasets.

In contrast, this study adopts a Requirement-Oriented Benchmarking (ROB) perspective to systematically evaluate seven neural network architectures—CNN, RNN, GNN, Autoencoder, Transformer, Neural Collaborative Filtering, and Siamese Networks—under a unified evaluation protocol. By analyzing performance across multiple datasets and system requirements, including accuracy, diversity, relational awareness, temporal dynamics, and computational efficiency, our work provides comparative insights that complement existing literature rather than proposing new recommendation models.
Table~\ref{tab:sota_summary} summarizes representative prior studies and highlights the lack of unified, multi-objective benchmarking across architectures. By addressing this gap, the present work offers a structured evaluation framework and empirical evidence that inform architecture selection and hybrid design strategies for modern recommendation systems.

\begin{table*}[h!]
\centering
\caption{Summary of State-of-the-Art Works in Recommendation Systems}
\label{tab:sota_summary}
\resizebox{\textwidth}{!}{%
\begin{tabular}{p{1.3cm} p{1.3cm} p{3.5cm} p{4.2cm} p{4.2cm} p{3cm}}
\toprule
\textbf{Ref} & \textbf{Year} & \textbf{Approach} & \textbf{Key Advantages} & \textbf{Main Limitations} & \textbf{Dataset(s)} \\
\midrule
\cite{25} & 2024 & ClusterFedMet & Enhances personalization in federated settings & Privacy and security aspects not fully explored & -- \\

\cite{26} & 2024 & Graph-based aesthetic assessment & Improves visual recommendation quality & Computationally intensive; requires high-quality images & Image datasets \\

\cite{27} & 2023 & Content-based filtering & Increases novelty and diversity & Risk of overspecialization & MovieLens 1M \\

\cite{28} & 2023 & Sentiment-based hybrid RS & Improves accuracy through sentiment analysis & High computational complexity & BeerAdvocate, Modcloth, Amazon \\

\cite{33} & 2023 & Spiking Neural Networks (SNNs) & Energy-efficient neuromorphic processing & Complex implementation; limited scalability & -- \\

\cite{34} & 2024 & NAS for SNNs & Optimized SNN architectures & Requires extensive computational resources & -- \\

\cite{35} & 2023 & HybridBranchNet & Scalable CNN-based recommendation & Sensitive to hyperparameter tuning & Image datasets \\

\cite{36} & 2024 & RS network connectivity analysis & Improves user navigation efficiency & Trade-off between depth and breadth & Steam, MovieLens, Amazon \\

\cite{37} & 2023 & Financial RS accuracy study & High predictive accuracy in financial domains & Limited generalizability & BRICS stock markets \\

\cite{43} & 2024 & LAP-SR & Enhances long-tail item exposure & Algorithmic complexity & E-commerce datasets \\

\cite{44} & 2024 & Teddy model & Joint modeling of trends and diversity & Dual-path architectural complexity & E-commerce datasets \\

\cite{45} & 2023 & DI-RAR & Captures implicit API requirements & High computational overhead & API datasets \\

\cite{46} & 2024 & Earthquake-aware RS & Improves public preparedness via diversity & Domain-specific applicability & EU seismic datasets \\

\cite{47} & 2023 & Sensitive-user RS & Balances influential and sensitive items & Scalability challenges & Social networks \\

\textbf{Our Approach} & 2025 & Requirement-Oriented benchmarking framework & Multi-objective evaluation; Practical guidance for designing modern recommendation systems & Lack LLM and computational evaluations & Retail Rocket, Amazon, Netflix \\
\bottomrule
\end{tabular}}
\end{table*}

\section{Method}

The methodological design of this study is guided by a ROB perspective, which structures evaluation around the core demands of modern recommendation systems. These requirements are operationalized through a unified evaluation protocol to ensure fair, consistent, and reproducible comparison across neural architectures. In this section, we describe the experimental methodology used to evaluate seven neural network architectures for item–item recommendation tasks. The study is conducted on three heterogeneous datasets, Amazon, Netflix Prize, and Retail Rocket E- commerce, each exhibiting distinct interaction characteristics. Raw data from these sources undergoes standardized preprocessing to ensure consistency, completeness, and suitability for machine learning workflows. To ensure a fair and systematic comparison, all models are evaluated under a unified, requirement-driven evaluation protocol aligned with the core demands of modern recommendation systems. Specifically, the evaluation considers predictive accuracy (R1), recommendation diversity (R2), relational awareness (R3), temporal modeling capability (R4), and computational efficiency (R5). These requirements are operationalized through consistent preprocessing, training procedures, evaluation metrics, and computational measurements across all architectures.

Each of the seven neural architectures: CNN, RNNs, GNNs, Autoencoders, Transformers, NCF, and Siamese Networks—is then applied to the datasets. The evaluation focuses on each model’s ability to balance precision and diversity in recommendations. Performance is assessed using precision, recall, diversity score, and computational efficiency, allowing a thorough comparison of the strengths and limitations of each architecture in realistic e- commerce scenarios.

\begin{figure*}[h!]
    \centering
    \includegraphics[width=1.05\linewidth]{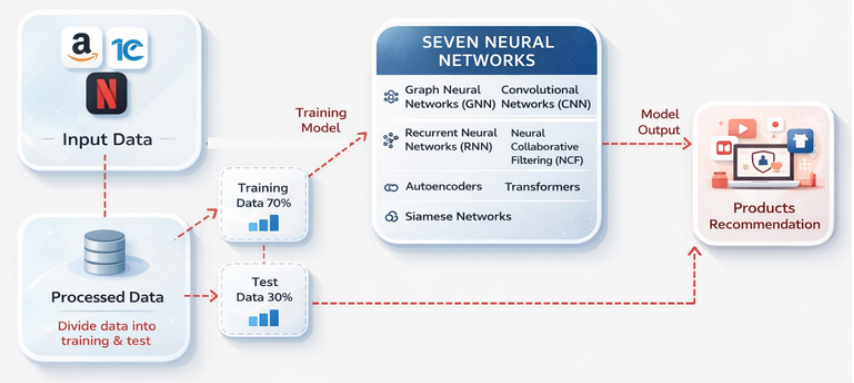}
    \caption{General workflow of the proposed approach.}
    \label{fig:workflow}
\end{figure*}

This comparative analysis aims to provide insights into which models best enhance recommendation quality, scalability, and adaptability. All data handling procedures strictly follow privacy-preserving principles and ethical standards. Figure~\ref{fig:workflow} presents the general workflow of the proposed approach.

\subsection{Data Preprocessing}

Before model implementation, an extensive preprocessing phase is conducted to prepare the datasets. This step is essential for ensuring data quality and maximizing the effectiveness of the neural architectures. The preprocessing pipeline includes missing-value handling, normalization, categorical encoding, and dimensionality reduction. The data is then divided into training and testing subsets. The overall procedure is summarized in Algorithm~\ref{alg:preprocessing}. For the GNN-based models, datasets are structurally converted into undirected graphs. Each node represents an item and contains a single attribute (type) to preserve privacy. Edges encode relationships such as cooccurrences or interactions. The graph construction process is detailed in Algorithm~\ref{alg:graph}. All datasets undergo identical preprocessing pipelines to ensure comparability across architectures and to eliminate confounding factors introduced by dataset-specific handling.

\begin{algorithm*}[t!]
\small
\caption{Data Preprocessing Pipeline}
\label{alg:preprocessing}

\textbf{Step 1: Input:} Raw\_Data (user interactions, item metadata, auxiliary fields).\\
\textbf{Step 2: Output:} Preprocessed\_Data ready for model training and evaluation.

\textbf{Step 3: Data Cleaning:}\\
\hspace*{1em}\textbf{3.1:} Remove corrupted or incomplete entries.\\
\hspace*{1em}\textbf{3.2:} Fill missing values using statistical or model-based imputation.\\
\hspace*{1em}\textbf{3.3:} Remove duplicate records.

\textbf{Step 4: Text Normalization (if applicable):}\\
\hspace*{1em}\textbf{4.1:} Convert all text to lowercase.\\
\hspace*{1em}\textbf{4.2:} Remove punctuation, symbols, and extra whitespace.\\
\hspace*{1em}\textbf{4.3:} Apply stemming or lemmatization.

\textbf{Step 5: Tokenization and Encoding:}\\
\hspace*{1em}\textbf{5.1:} Tokenize text fields.\\
\hspace*{1em}\textbf{5.2:} Encode categorical variables (Label Encoding, One-Hot, TF-IDF, or embeddings).\\
\hspace*{1em}\textbf{5.3:} Build word/item embeddings if required.

\textbf{Step 6: Numerical Feature Scaling:}\\
\hspace*{1em}\textbf{6.1:} Apply MinMaxScaler, StandardScaler, or normalization to a common range.

\textbf{Step 7: Feature Optimization:}\\
\hspace*{1em}\textbf{7.1:} Perform feature selection (filter, wrapper, or embedded methods).\\
\hspace*{1em}\textbf{7.2:} Optionally apply dimensionality reduction (PCA, SVD, Autoencoders).

\textbf{Step 8: Finalization:}\\
\hspace*{1em}\textbf{8.1:} Merge cleaned, encoded, and scaled variables.\\
\hspace*{1em}\textbf{8.2:} Output Preprocessed\_Data.
\end{algorithm*}

\begin{algorithm*}[t!]
\small
\caption{Graph Construction for GNN-Based Recommendation}
\label{alg:graph}

\textbf{Step 1: Input:} Preprocessed\_Data containing item interactions, co-occurrences, and item attributes.\\
\textbf{Step 2: Output:} Training graph $G$ and testing graph $G'$.

\textbf{Step 3: Dataset Splitting:}\\
\hspace*{1em}\textbf{3.1:} Split Preprocessed\_Data into 70\% Train\_Data and 30\% Test\_Data.\\
\hspace*{1em}\textbf{3.2:} Initialize two empty graphs $G$ (training) and $G'$ (testing).

\textbf{Step 4: Node Construction:}\\
\hspace*{1em}\textbf{4.1:} For each item in Train\_Data:\\
\hspace*{2em}Insert a node with attribute: type $\rightarrow$ \texttt{category}.\\
\hspace*{1em}\textbf{4.2:} Repeat the same process for Test\_Data to build $G'$.

\textbf{Step 5: Edge Construction:}\\
\hspace*{1em}\textbf{5.1:} For each user session or order in Train\_Data:\\
\hspace*{2em}Add edges between items purchased/viewed together (co-occurrence).\\
\hspace*{2em}Assign edge weight $w_{ij}$ proportional to co-occurrence frequency.\\
\hspace*{1em}\textbf{5.2:} Apply the same procedure to Test\_Data for graph $G'$.

\textbf{Step 6: Final Output:}\\
\hspace*{1em}\textbf{6.1:} Return training graph $G$ and testing graph $G'$ for GNN model usage.
\end{algorithm*}

\subsection{Model Selection}

The seven neural architectures were selected to provide a comprehensive evaluation of different deep learning paradigms used in item–item recommendation. Each architecture contributes unique strengths:

\begin{itemize}
    \item \textbf{CNNs:} Capture hierarchical patterns in interaction matrices through spatial feature extraction.
    \item \textbf{RNNs:} Model sequential dependencies, suitable for temporal user behavior.
    \item \textbf{GNNs:} Naturally process graph-structured data representing item relationships.
    \item \textbf{Autoencoders:} Learn compact latent representations, useful for dimensionality reduction.
    \item \textbf{Transformers:} Capture long-range dependencies using self-attention mechanisms.
    \item \textbf{NCF:} Learn non-linear user–item interactions by combining MLPs and generalized matrix factorization.
    \item \textbf{Siamese Networks:} Measure item similarity by comparing item pairs, enhancing diversity.
\end{itemize}

This selection ensures a balanced comparison of accuracy, diversity, scalability, and representational capacity.

\subsection{Requirement Oriented Evaluation Metrics}
To align the evaluation with the core requirements of modern recommendation systems, each system requirement is operationalized through established quantitative metrics. Predictive accuracy (R1) is measured using precision, recall, and F1-score to assess the relevance of recommended items. Recommendation diversity (R2) is quantified using a diversity score that reflects the heterogeneity of recommended item sets. Relational awareness (R3) and temporal dynamics (R4) are evaluated indirectly through model performance on datasets that explicitly encode graph-based relationships and sequential user behavior, respectively. Computational efficiency (R5) is assessed by measuring training time and inference cost under consistent hardware and implementation settings. This requirement-oriented evaluation framework enables a fair and comprehensive comparison across architectures with different modeling inductive biases. Within the requirement- oriented benchmarking approach adopted in this study, each system requirement is evaluated using established quantitative metrics. Table~\ref{tab:modern_requirements} summarizes the correspondence between requirements and evaluation metrics.

\begin{table*}[h!]
\centering
\caption{Modern Recommendation System Requirements}
\label{tab:modern_requirements}
\resizebox{\textwidth}{!}{%
\begin{tabular}{p{3cm} p{6.5cm} p{6.5cm}}
\toprule
\textbf{Requirement ID} & \textbf{System Requirement} & \textbf{Evaluation Metrics / Indicators} \\
\midrule
R1 & Predictive Accuracy &
Accuracy@K, Recall, F1-score, NDCG \\

R2 & Recommendation Diversity &
Intra-list Diversity (ILD), Coverage, Novelty \\

R3 & Relational Awareness &
Performance on graph-based datasets, Graph Sensitivity Analysis \\

R4 & Temporal Dynamics &
Sequential Recall@K, Time-aware NDCG, Performance on temporal datasets \\

R5 & Computational Efficiency &
Training Time (s), Inference Latency (ms), Memory Consumption (GB) \\
\bottomrule
\end{tabular}
}
\end{table*}

\subsection{Model Implementation}
The selected neural architectures represent complementary modeling paradigms commonly used in recommendation systems, each corresponding to distinct system requirements such as relational modeling, temporal dependency capture, representation learning, and diversity promotion. Table~\ref{tab:implementation} summarizes the implementation details for each model, including tools, configurations, and training parameters, enabling reproducibility.

\begin{table*}[h!]
\centering
\caption{Implementation Overview of the Evaluated Recommendation Models}
\label{tab:implementation}
\resizebox{\textwidth}{!}{%
\begin{tabular}{p{3.2cm} p{3.5cm} p{6.5cm} p{4.5cm}}
\toprule
\textbf{Model Type} & \textbf{Software / Tools} & \textbf{Architecture / Configuration} & \textbf{Training Parameters} \\
\midrule
CNNs & TensorFlow, Keras &
2D convolutions with ReLU activation, max-pooling layers, and dropout regularization &
Learning rate: 0.001; Batch size: 128; Epochs: 30 \\

RNNs & PyTorch &
LSTM-based recurrent units with dropout rate of 0.5 &
Sequence length: 10; Learning rate: 0.01; Batch size: 64; Epochs: 50 \\

GNNs & DGL, PyTorch &
Item--item graph neural network with message passing &
Learning rate: 0.005; Epochs: 40 \\

Autoencoders & TensorFlow, Keras &
Symmetric dense encoder--decoder architecture with sigmoid output &
Learning rate: 0.001; Batch size: 256; Epochs: 50 \\

Transformers & HuggingFace Transformers &
BERT-like encoder with positional encoding and multi-head self-attention &
Learning rate: 0.0001; Batch size: 32; Epochs: 20 \\

NCF & TensorFlow, Keras &
Hybrid model combining MLP and generalized matrix factorization &
Learning rate: 0.0005; Batch size: 128; Epochs: 30 \\

Siamese Networks & TensorFlow, Keras &
Twin subnetworks with shared weights for similarity learning &
Learning rate: 0.0005; Batch size: 64; Epochs: 35 \\
\bottomrule
\end{tabular}}
\end{table*}

\section{Experimentation and Results}

\subsection*{Research Questions}
To evaluate neural architectures under the ROB framework, the following research questions are defined:

\begin{itemize}
    \item \textbf{RQ1 (R1 – Predictive Accuracy):} How do different neural network architectures perform in terms of accuracy across heterogeneous recommendation datasets?
    \item \textbf{RQ2 (R1 – Accuracy Robustness):} Which models provide the highest recall and F1-score, and how consistent are these performances across datasets?
    \item \textbf{RQ3 (R1/R4 – Ranking Robustness):} How does top-$k$ recommendation performance degrade as the recommendation list expands, and which architectures maintain stable performance at larger $k$ values?
    \item \textbf{RQ4 (R2 – Recommendation Diversity):} Which neural network architecture provides the greatest intra-list diversity in top-$k$ recommendations?
    \item \textbf{RQ5 (R1–R2 Trade-off):} How do predictive accuracy and recommendation diversity interact within each model, and which architectures achieve the most favorable balance between relevance and diversity?
\end{itemize}
\noindent\textit{Note:} Requirements R3 (Relational Awareness) and R5 (Computational Efficiency) are evaluated through comparative performance analysis on graph-structured datasets and explicit measurements of training time, inference latency, and memory consumption, respectively, as reported in the subsequent experimental and results sections.

These research questions are addressed systematically in the following subsections, with results interpreted in relation to the core system requirements defined by the ROB framework.
\subsection{Dataset and Applications Contexts }
Three real-world datasets from distinct application domains are used to evaluate the generality and robustness of the proposed benchmark. All datasets are preprocessed following the procedure described in Section III to ensure consistency across experiments.

\textbf{Retail Rocket Dataset}: This dataset contains category hierarchies, item attributes, and user interaction logs, capturing consumer behavior such as product views and checkouts. It reflects typical e-commerce interaction patterns.

\textbf{Amazon Product Dataset}: Sourced from Amazon, this dataset includes product metadata and user-generated reviews, providing rich signals of item popularity and user–item interactions.

\textbf{Netflix Prize Dataset}: This dataset consists of timestamped user ratings for movies and is well suited for analyzing temporal user preferences and recommendation performance in streaming platforms.

\subsection{Evaluation Metrics}
To quantitatively assess model performance under the ROB framework, we employ widely adopted evaluation metrics that capture predictive accuracy and recommendation diversity. These metrics are formally defined below and are used consistently across all experiments.
\textbf{Accuracy:}
\begin{equation}
\text{Accuracy} = \frac{TP + TN}{TP + TN + FP + FN}
\end{equation}
where \(TP\) is the number of true positives, \(TN\) is the number of true negatives, \(FP\) is the number of false positives, and \(FN\) is the number of false negatives.

\textbf{Accuracy@k:}
\begin{equation}
\text{Accuracy@k} = \frac{\text{Number of relevant items in top-$k$}}{k}
\end{equation}
where \(k\) is the number of top recommended items and “relevant items” are those that match the user’s actual preferences.

\textbf{Recall:}
\begin{equation}
\text{Recall} = \frac{TP}{TP + FN}
\end{equation}
where \(TP\) is the number of true positives and \(FN\) is the number of false negatives.

\textbf{F1-Score:}
\begin{equation}
\text{F1-Score} = \frac{2 \times \text{Precision} \times \text{Recall}}{\text{Precision} + \text{Recall}}
\end{equation}
where \(\text{Precision} = \frac{TP}{TP+FP}\) and \(\text{Recall} = \frac{TP}{TP+FN}\).
\\
\textbf{Intra-list Diversity@k (ILD@k):}
\begin{equation}
\text{ILD@k} = 1 - \frac{1}{k (k-1)} 
\sum_{i=1}^{k} 
\sum_{\substack{j=1 \\ j \neq i}}^{k} 
\text{similarity}(item_i, item_j)
\end{equation}
where \(k\) is the number of recommended items considered, and \(\text{similarity}(item_i, item_j)\) measures how similar two items are (e.g., based on feature embeddings or user ratings).

\subsection{Justification of Model Parameters}

Model configurations were selected through iterative tuning to ensure stable convergence and competitive performance across datasets, while maintaining consistency with the unified evaluation protocol. The following summarizes the architectural choices and key parameters for each model.

\textbf{CNNs:} Use 2D convolutions with ReLU for nonlinear learning, dropout for regularization, and max-pooling to reduce dimensionality.

\textbf{RNNs:} LSTM units were chosen for capturing temporal patterns. A dropout of 0.5 prevents overfitting.

\textbf{GNNs:} Two graph convolution layers enable learning representations based on user-item connections. A learning rate of 0.005 ensures stable convergence.

\textbf{Autoencoders:} Symmetric encoder-decoder architecture captures latent representations, with sigmoid normalization in the output.

\textbf{Transformers:} Multi-head attention enables modeling long-range dependencies; positional encoding preserves sequence order.

\textbf{NCF:} Combines matrix factorization with MLP to model nonlinear user-item interactions.

\textbf{Siamese Networks:} Designed for similarity learning, providing strong diversity in item-to-item recommendations.

\subsection{Predictive Accuracy and Robustness Across Datasets}
Tables~\ref{tab:retail_metrics},~\ref{tab:amazon_metrics},~\ref{tab:netflix_metrics}, visualized in Figures~\ref{fig:metrics_5_1_5_02},~\ref{fig:metrics_all}, summarize predictive accuracy, recall, and F1-score for all evaluated architectures across the Retail Rocket, Amazon, and Netflix Prize datasets. Under the ROB framework, these results correspond to Requirement R1 (Predictive Accuracy) and provide insight into the robustness of each model across heterogeneous application contexts. GNNs consistently achieve the highest performance on the Retail Rocket and Amazon datasets, reflecting their ability to leverage relational item– item structures common in e-commerce environments. In contrast, RNN exhibit the strongest performance on the Netflix Prize dataset, highlighting their effectiveness in modeling sequential user behavior in temporally driven recommendation scenarios. Transformer-based models also demonstrate competitive accuracy across datasets, particularly in leading positions where long-range dependencies are prominent. Autoencoder-based approaches show comparatively lower performance across all datasets, suggesting limitations in capturing complex interaction patterns when used in
isolation. Overall, these findings indicate that no single architecture universally dominates across datasets, reinforcing the importance of selecting models in accordance with specific system requirements.

\begin{table}[h!]
\centering
\caption{Performance metrics on Retail Rocket E-commerce dataset}
\label{tab:retail_metrics}
\begin{tabular}{p{1.5cm}p{1.6cm}p{1.4cm}p{1.8cm}p{0.6cm}}

\toprule
\textbf{Model} & \textbf{Accuracy(\%)} & \textbf{Recall(\%)} & \textbf{F1-Score(\%)} & \textbf{TT(s)} \\
\midrule
CNN & 80 & 83 & 80.5 & 420 \\
RNN & 83 & 86 & 83.5 & 680 \\
\textbf{GNN} & \textbf{88} & \textbf{91} & \textbf{88.5} & 910 \\
Autoencoder & 73 & 76 & 73.5 & 360 \\
Transformer & 86 & 88 & 86.5 & 1150 \\
NCF & 82 & 81 & 82.5 & 290 \\
Siamese & 81 & 83 & 81.5 & 520 \\
\bottomrule
\end{tabular}
\end{table}

\begin{table}[h!]
\centering
\caption{Performance metrics on Amazon dataset}
\label{tab:amazon_metrics}
\begin{tabular}{p{1.5cm}p{1.6cm}p{1.4cm}p{1.8cm}p{0.6cm}}
\toprule
\textbf{Model} & \textbf{Accuracy(\%)} & \textbf{Recall(\%)} & \textbf{F1-Score(\%)} & \textbf{TT(s)} \\
\midrule
CNN & 78 & 79 & 77.5 & 2\,850 \\
RNN & 82 & 85 & 82.5 & 4\,200 \\
\textbf{GNN} & \textbf{90} & \textbf{93} & \textbf{90.5} & 5\,800 \\
Autoencoder & 70 & 73 & 70.5 & 2\,100 \\
Transformer & 88 & 90 & 88 & 7\,900 \\
NCF & 85 & 84 & 85.5 & 1\,750 \\
Siamese & 84 & 86 & 84.5 & 3\,100 \\
\bottomrule
\end{tabular}
\end{table}

\begin{table}[h!]
\centering
\caption{Performance metrics on Netflix Prize dataset}
\label{tab:netflix_metrics}
\begin{tabular}{p{1.5cm}p{1.6cm}p{1.4cm}p{1.8cm}p{0.6cm}}
\toprule
\textbf{Model} & \textbf{Accuracy(\%)} & \textbf{Recall(\%)} & \textbf{F1-Score(\%)} & \textbf{TT(s)} \\
\midrule
CNN & 75 & 78 & 75.5 & 1\,120 \\
\textbf{RNN} & \textbf{89} & \textbf{89} & \textbf{89.5} & 1\,950 \\
GNN & 85 & 87 & 85.5 & 2\,600 \\
Autoencoder & 72 & 75 & 72.5 & 980 \\
Transformer & 83 & 85 & 83.5 & 3\,400 \\
NCF & 80 & 79 & 80 & 820 \\
Siamese & 78 & 80 & 78.5 & 1\,480 \\
\bottomrule
\end{tabular}
\end{table}


\begin{table*}[htbp!]
\centering
\caption{Performance of top-$k$ recommendations across all datasets for all models}
\label{tab:all_models}

\begin{minipage}{0.49\textwidth}
\centering
\renewcommand{\arraystretch}{1.1}
\setlength{\tabcolsep}{4pt}
\subcaption*{(a) CNN}
\begin{tabular}{c cc cc cc}
\toprule
\textbf{Top-$k$} 
& \multicolumn{2}{c}{\textbf{Retail}} 
& \multicolumn{2}{c}{\textbf{Amazon}} 
& \multicolumn{2}{c}{\textbf{Netflix}} \\
\cmidrule(lr){2-3} \cmidrule(lr){4-5} \cmidrule(lr){6-7}
& Acc & ILD & Acc & ILD & Acc & ILD \\
\midrule
1  & 86.12 & 0.00  & 84.76 & 0.00  & 82.66 & 0.00 \\
2  & 81.46 & 21.34 & 80.30 & 21.75 & 78.22 & 22.37 \\
3  & 79.15 & 24.72 & 78.08 & 25.18 & 76.04 & 25.88 \\
4  & 77.98 & 29.49 & 76.43 & 30.04 & 74.64 & 31.14 \\
5  & 75.85 & 37.27 & 74.36 & 37.75 & 72.46 & 39.11 \\
6  & 71.12 & 42.72 & 70.15 & 43.42 & 67.86 & 45.12 \\
7  & 64.68 & 47.45 & 63.61 & 48.14 & 61.83 & 49.63 \\
8  & 61.50 & 55.10 & 60.46 & 56.06 & 59.14 & 57.46 \\
9  & 58.36 & 61.71 & 57.54 & 62.88 & 55.92 & 64.97 \\
10 & 56.21 & 66.37 & 55.09 & 67.23 & 53.40 & 69.40 \\
\bottomrule
\end{tabular}

\end{minipage}
\hfill
\begin{minipage}{0.49\textwidth}
\centering
\renewcommand{\arraystretch}{1.1}
\setlength{\tabcolsep}{4pt}
\subcaption*{(b) RNN}
\begin{tabular}{c cc cc cc}
\toprule
\textbf{Top-$k$} 
& \multicolumn{2}{c}{\textbf{Retail}} 
& \multicolumn{2}{c}{\textbf{Amazon}} 
& \multicolumn{2}{c}{\textbf{Netflix}} \\
\cmidrule(lr){2-3} \cmidrule(lr){4-5} \cmidrule(lr){6-7}
& Acc & ILD & Acc & ILD & Acc & ILD \\
\midrule
1  & 89.43 & 0.00  & 88.48 & 0.00  & 93.41 & 0.00 \\
2  & 86.65 & 20.64 & 85.61 & 20.92 & 89.49 & 21.33 \\
3  & 80.51 & 25.62 & 79.63 & 25.98 & 83.90 & 26.74 \\
4  & 77.98 & 29.39 & 76.89 & 29.75 & 81.31 & 30.71 \\
5  & 74.45 & 32.83 & 73.70 & 33.19 & 77.18 & 34.30 \\
6  & 71.12 & 38.27 & 70.39 & 38.82 & 73.91 & 39.68 \\
7  & 65.78 & 45.45 & 64.39 & 46.12 & 68.28 & 47.27 \\
8  & 61.90 & 49.10 & 61.19 & 49.69 & 64.12 & 51.12 \\
9  & 58.72 & 54.85 & 57.94 & 55.59 & 61.08 & 56.86 \\
10 & 57.12 & 60.97 & 56.47 & 61.75 & 59.59 & 63.40 \\
\bottomrule
\end{tabular}

\end{minipage}

\begin{minipage}{0.49\textwidth}
\centering
\renewcommand{\arraystretch}{1.1}
\setlength{\tabcolsep}{4pt}
\subcaption*{(c) GNN}
\begin{tabular}{c cc cc cc}
\toprule
\textbf{Top-$k$} 
& \multicolumn{2}{c}{\textbf{Retail}} 
& \multicolumn{2}{c}{\textbf{Amazon}} 
& \multicolumn{2}{c}{\textbf{Netflix}} \\
\cmidrule(lr){2-3} \cmidrule(lr){4-5} \cmidrule(lr){6-7}
& Acc & ILD & Acc & ILD & Acc & ILD \\
\midrule
1  & 95.63 & 0.00  & 98.47 & 0.00  & 88.39 & 0.00 \\
2  & 92.76 & 19.34 & 96.12 & 7.37  & 85.83 & 32.43 \\
3  & 91.11 & 23.62 & 94.38 & 12.49 & 82.59 & 36.73 \\
4  & 87.98 & 28.19 & 91.93 & 18.25 & 78.29 & 40.19 \\
5  & 84.45 & 33.83 & 86.43 & 30.92 & 74.30 & 47.32 \\
6  & 79.12 & 37.27 & 81.09 & 38.12 & 68.34 & 52.49 \\
7  & 77.28 & 39.45 & 77.32 & 44.16 & 61.95 & 55.91 \\
8  & 74.90 & 42.10 & 74.31 & 47.59 & 55.12 & 62.03 \\
9  & 69.42 & 52.85 & 70.98 & 56.42 & 49.29 & 67.30 \\
10 & 66.12 & 56.97 & 67.04 & 59.31 & 44.21 & 71.23 \\
\bottomrule
\end{tabular}

\end{minipage}
\hfill
\begin{minipage}{0.49\textwidth}
\centering
\renewcommand{\arraystretch}{1.1}
\setlength{\tabcolsep}{4pt}
\subcaption*{(d) Autoencoder}
\begin{tabular}{c cc cc cc}
\toprule
\textbf{Top-$k$} 
& \multicolumn{2}{c}{\textbf{Retail}} 
& \multicolumn{2}{c}{\textbf{Amazon}} 
& \multicolumn{2}{c}{\textbf{Netflix}} \\
\cmidrule(lr){2-3} \cmidrule(lr){4-5} \cmidrule(lr){6-7}
& Acc & ILD & Acc & ILD & Acc & ILD \\
\midrule
1  & 80.15 & 0.00  & 76.14 & 0.00  & 76.14 & 0.00 \\
2  & 75.41 & 20.54 & 73.15 & 21.20 & 73.15 & 21.88 \\
3  & 73.23 & 25.13 & 71.03 & 25.98 & 71.03 & 26.86 \\
4  & 72.54 & 30.05 & 70.36 & 31.13 & 69.09 & 32.25 \\
5  & 71.23 & 36.13 & 69.09 & 37.50 & 62.74 & 38.93 \\
6  & 65.91 & 39.88 & 62.74 & 41.48 & 56.91 & 43.14 \\
7  & 59.82 & 42.29 & 56.91 & 44.07 & 51.34 & 45.92 \\
8  & 57.72 & 56.87 & 55.99 & 59.37 & 49.32 & 61.98 \\
9  & 54.16 & 59.31 & 52.34 & 62.04 & 48.32 & 64.89 \\
10 & 52.67 & 62.40 & 51.09 & 65.40 & 46.51 & 68.54 \\
\bottomrule
\end{tabular}

\end{minipage}

\vspace{0.3cm}

\begin{minipage}{0.49\textwidth}
\centering
\renewcommand{\arraystretch}{1.1}
\setlength{\tabcolsep}{4pt}
\subcaption*{(e) Transformer}
\begin{tabular}{c cc cc cc}
\toprule
\textbf{Top-$k$} 
& \multicolumn{2}{c}{\textbf{Retail}} 
& \multicolumn{2}{c}{\textbf{Amazon}} 
& \multicolumn{2}{c}{\textbf{Netflix}} \\
\cmidrule(lr){2-3} \cmidrule(lr){4-5} \cmidrule(lr){6-7}
& Acc & ILD & Acc & ILD & Acc & ILD \\
\midrule
1  & 90.43 & 0.00  & 97.28 & 0.00  & 85.39 & 0.00 \\
2  & 88.63 & 19.21 & 94.12 & 9.97  & 83.83 & 28.43 \\
3  & 83.51 & 24.67 & 93.36 & 11.59 & 79.59 & 31.73 \\
4  & 79.98 & 28.49 & 90.93 & 17.25 & 75.29 & 35.19 \\
5  & 77.32 & 31.43 & 87.13 & 28.42 & 71.30 & 39.32 \\
6  & 73.12 & 36.27 & 83.09 & 35.13 & 68.34 & 46.49 \\
7  & 67.58 & 44.47 & 80.42 & 41.16 & 63.95 & 49.91 \\
8  & 63.90 & 47.10 & 77.33 & 45.55 & 58.12 & 55.03 \\
9  & 60.82 & 52.15 & 74.98 & 52.42 & 51.29 & 59.30 \\
10 & 58.12 & 58.47 & 71.24 & 55.41 & 49.21 & 64.23 \\
\bottomrule
\end{tabular}

\end{minipage}
\hfill
\begin{minipage}{0.49\textwidth}
\centering
\renewcommand{\arraystretch}{1.1}
\setlength{\tabcolsep}{4pt}
\subcaption*{(f) NCF}
\begin{tabular}{c cc cc cc}
\toprule
\textbf{Top-$k$} 
& \multicolumn{2}{c}{\textbf{Retail}} 
& \multicolumn{2}{c}{\textbf{Amazon}} 
& \multicolumn{2}{c}{\textbf{Netflix}} \\
\cmidrule(lr){2-3} \cmidrule(lr){4-5} \cmidrule(lr){6-7}
& Acc & ILD & Acc & ILD & Acc & ILD \\
\midrule
1  & 89.31 & 0.00  & 96.18 & 0.00  & 83.23 & 0.00 \\
2  & 85.95 & 16.64 & 93.12 & 12.37 & 81.93 & 27.43 \\
3  & 82.51 & 19.62 & 91.36 & 15.59 & 77.59 & 30.73 \\
4  & 79.98 & 21.39 & 88.93 & 19.25 & 73.90 & 33.19 \\
5  & 72.45 & 24.83 & 85.13 & 23.82 & 70.20 & 37.32 \\
6  & 69.12 & 28.27 & 81.09 & 29.13 & 67.14 & 43.79 \\
7  & 67.78 & 35.45 & 77.42 & 36.66 & 62.85 & 47.41 \\
8  & 64.90 & 39.10 & 75.33 & 41.55 & 57.22 & 53.33 \\
9  & 59.72 & 44.85 & 71.98 & 46.32 & 50.59 & 58.20 \\
10 & 58.12 & 50.97 & 69.24 & 52.41 & 48.31 & 63.23 \\
\bottomrule
\end{tabular}

\end{minipage}

\vspace{0.3cm}

\begin{minipage}{0.49\textwidth}
\centering
\renewcommand{\arraystretch}{1.1}
\setlength{\tabcolsep}{4pt}
\subcaption*{(g) Siamese Networks}
\begin{tabular}{c cc cc cc}
\toprule
\textbf{Top-$k$} 
& \multicolumn{2}{c}{\textbf{Retail}} 
& \multicolumn{2}{c}{\textbf{Amazon}} 
& \multicolumn{2}{c}{\textbf{Netflix}} \\
\cmidrule(lr){2-3} \cmidrule(lr){4-5} \cmidrule(lr){6-7}
& Acc & ILD & Acc & ILD & Acc & ILD \\
\midrule
1  & 87.21 & 0.00  & 94.32 & 0.00  & 84.26 & 0.00 \\
2  & 82.29 & 21.74 & 92.15 & 17.37 & 80.50 & 21.71 \\
3  & 80.75 & 24.12 & 90.56 & 21.59 & 78.18 & 25.12 \\
4  & 78.91 & 29.42 & 87.13 & 25.25 & 76.83 & 30.48 \\
5  & 76.89 & 37.72 & 82.53 & 28.82 & 74.26 & 37.57 \\
6  & 72.13 & 42.25 & 80.29 & 32.13 & 70.25 & 43.24 \\
7  & 65.98 & 47.41 & 74.82 & 36.66 & 63.11 & 48.41 \\
8  & 62.10 & 55.19 & 72.13 & 44.55 & 60.96 & 56.60 \\
9  & 57.66 & 61.75 & 68.78 & 48.32 & 57.04 & 62.82 \\
10 & 56.81 & 66.32 & 65.44 & 55.41 & 55.29 & 67.32 \\
\bottomrule
\end{tabular}

\end{minipage}

\end{table*}


\subsection{Top-k Performance Degradation and Diversity Trade-offs}
\begin{figure*}[t!]
\centering

\begin{subfigure}{\textwidth}
    \centering
    \includegraphics[
        width=\textwidth,
        height=0.95\textheight,
        keepaspectratio
    ]{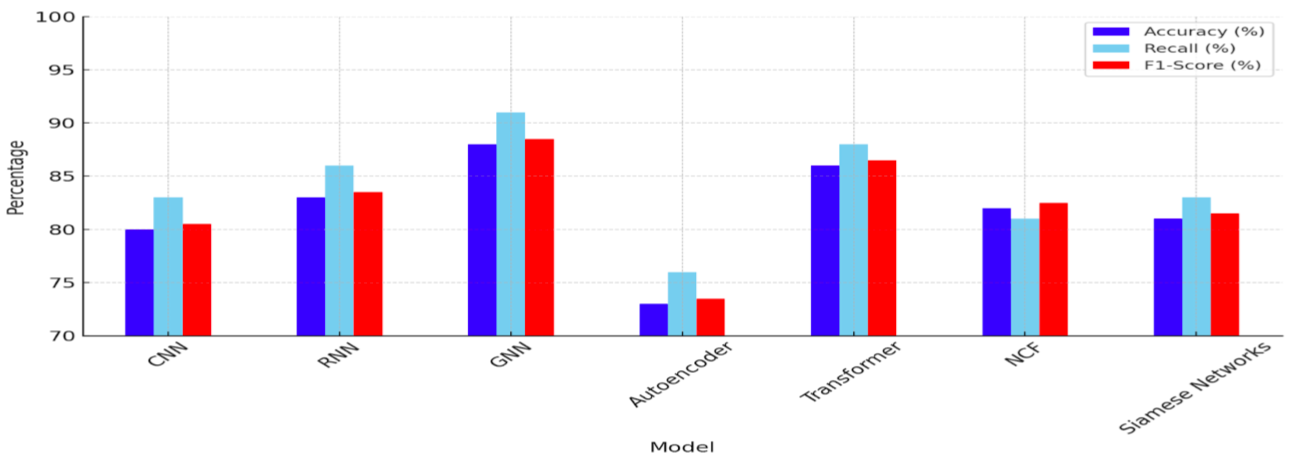}
    \caption{Retail Rocket E-commerce dataset}
\end{subfigure}

\vspace{0.5cm}

\begin{subfigure}{\textwidth}
    \centering
    \includegraphics[
        width=\textwidth,
        height=0.95\textheight,
        keepaspectratio
    ]{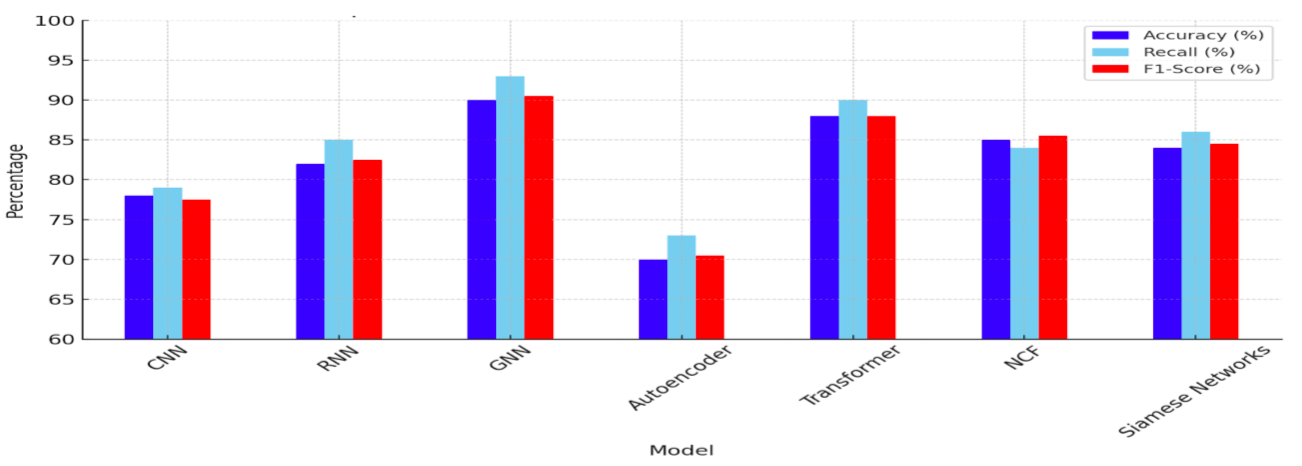}
    \caption{Amazon dataset}
\end{subfigure}

\vspace{0.5cm}

\begin{subfigure}{\textwidth}
    \centering
    \includegraphics[
        width=\textwidth,
        height=0.95\textheight,
        keepaspectratio
    ]{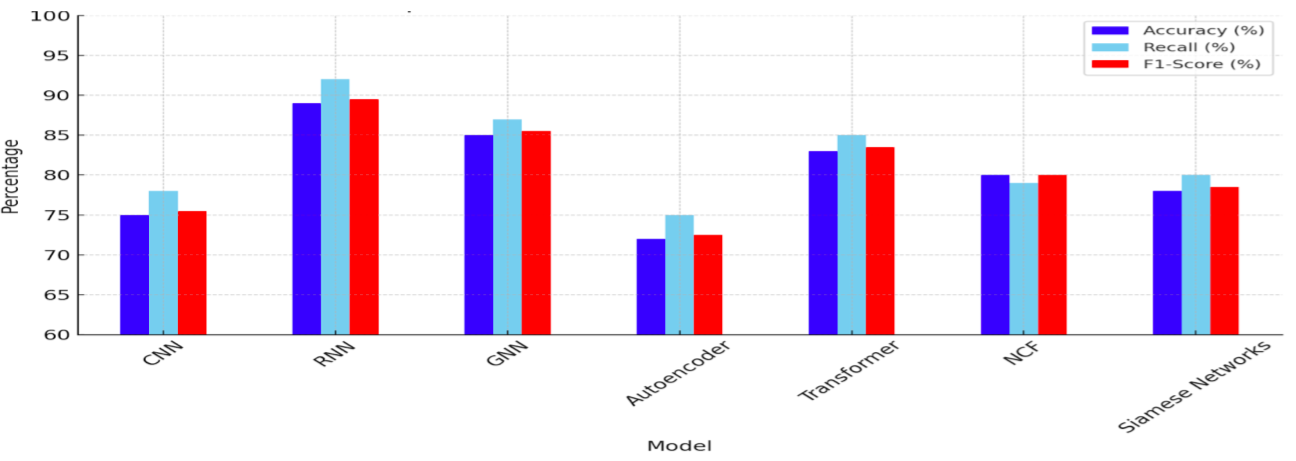}
    \caption{Netflix Prize dataset}
\end{subfigure}

\caption{Performance metrics visualization across three datasets:
(a) Retail Rocket E-commerce,
(b) Amazon,
and (c) Netflix Prize.}
\label{fig:metrics_all}
\end{figure*}

Table~\ref{tab:all_models} and Figure~\ref{fig:metrics_all} illustrate the evolution of predictive accuracy and intra-list diversity as the recommendation list size increases from top-1 to top-10. Under the ROB framework, these results jointly reflect Requirement R1 (predictive accuracy), Requirement R2 (recommendation diversity), and Requirement R4 (ranking robustness). Across all architectures and datasets, accuracy generally decreases as k increases, indicating the expected trade-off between recommendation breadth and relevance. In contrast, intra-list diversity consistently increases with larger k values, reflecting broader item exposure. GNN and RNN exhibit comparatively slower accuracy degradation while maintaining moderate diversity growth, suggesting greater robustness in expanding recommendation scenarios. Siamese Networks demonstrate the strongest diversity gains across datasets, particularly at larger k values, albeit with a more pronounced decline in accuracy. Autoencoder-based approaches show higher initial accuracy in small k settings but experience sharper degradation as k increases, indicating limited robustness in extended recommendation lists. These observations highlight that different architectures prioritize distinct objectives, reinforcing the multi-objective nature of modern recommendation systems.

\begin{figure*}[htbp!]
\centering

\begin{minipage}{0.49\textwidth}
    \centering
    \includegraphics[
  width=\linewidth,
  height=0.5\textheight,
  keepaspectratio
]{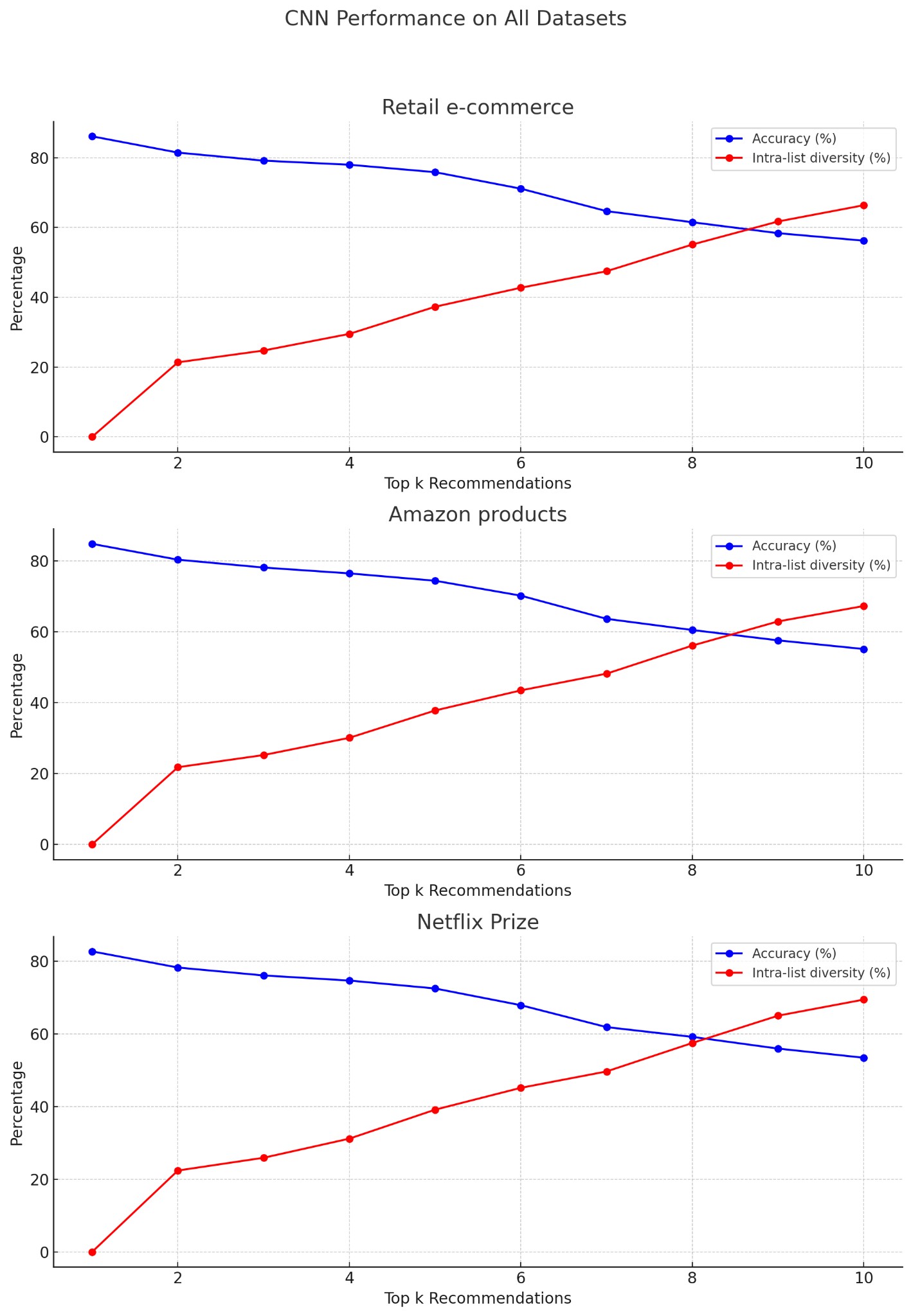}

\end{minipage}
\hfill
\begin{minipage}{0.49\textwidth}
    \centering
    \includegraphics[
  width=\linewidth,
  height=0.5\textheight,
  keepaspectratio
]{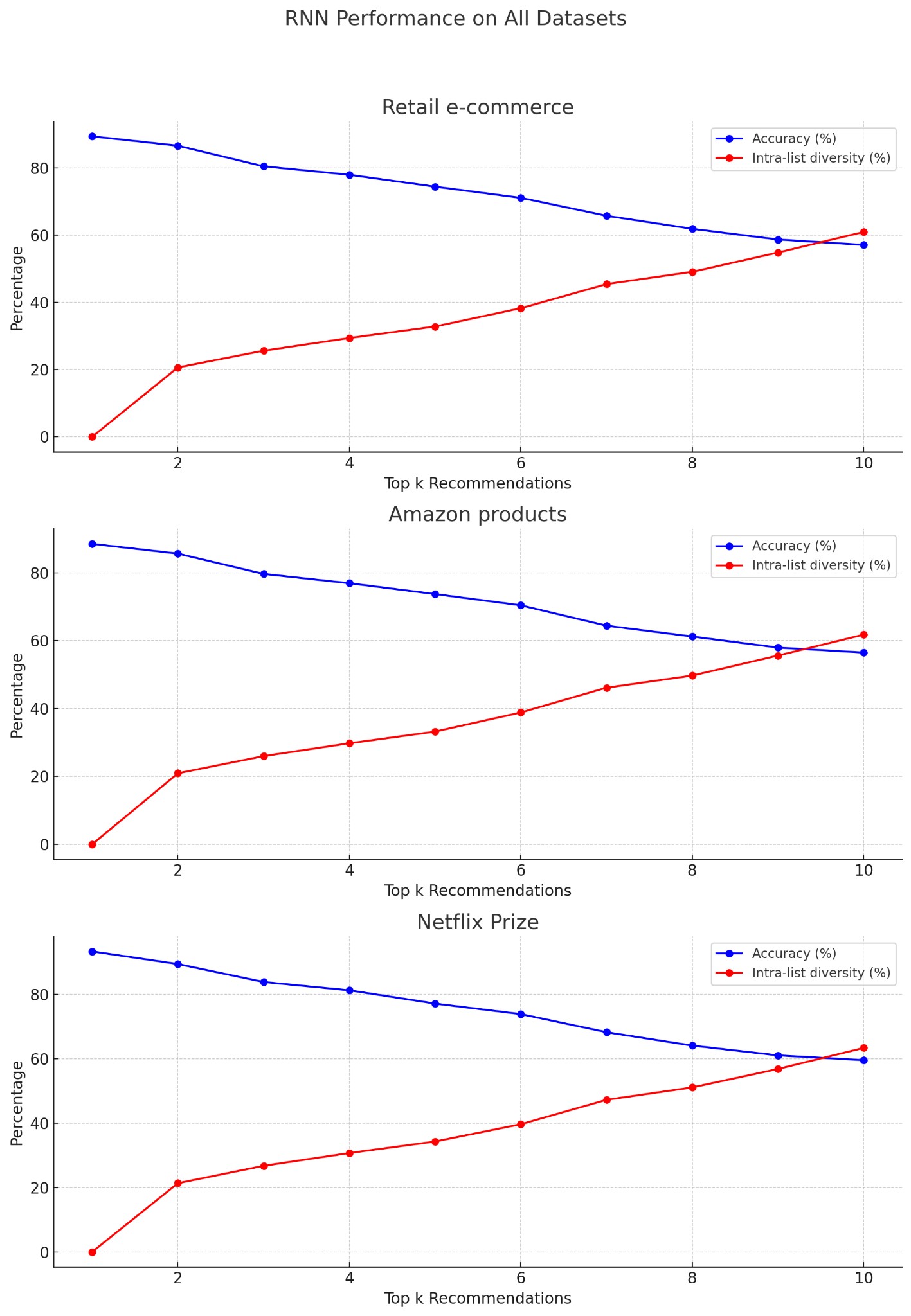}

\end{minipage}

\vspace{0.15cm}

\begin{minipage}{0.49\textwidth}
    \centering
    \includegraphics[
  width=\linewidth,
  height=0.5\textheight,
  keepaspectratio
]{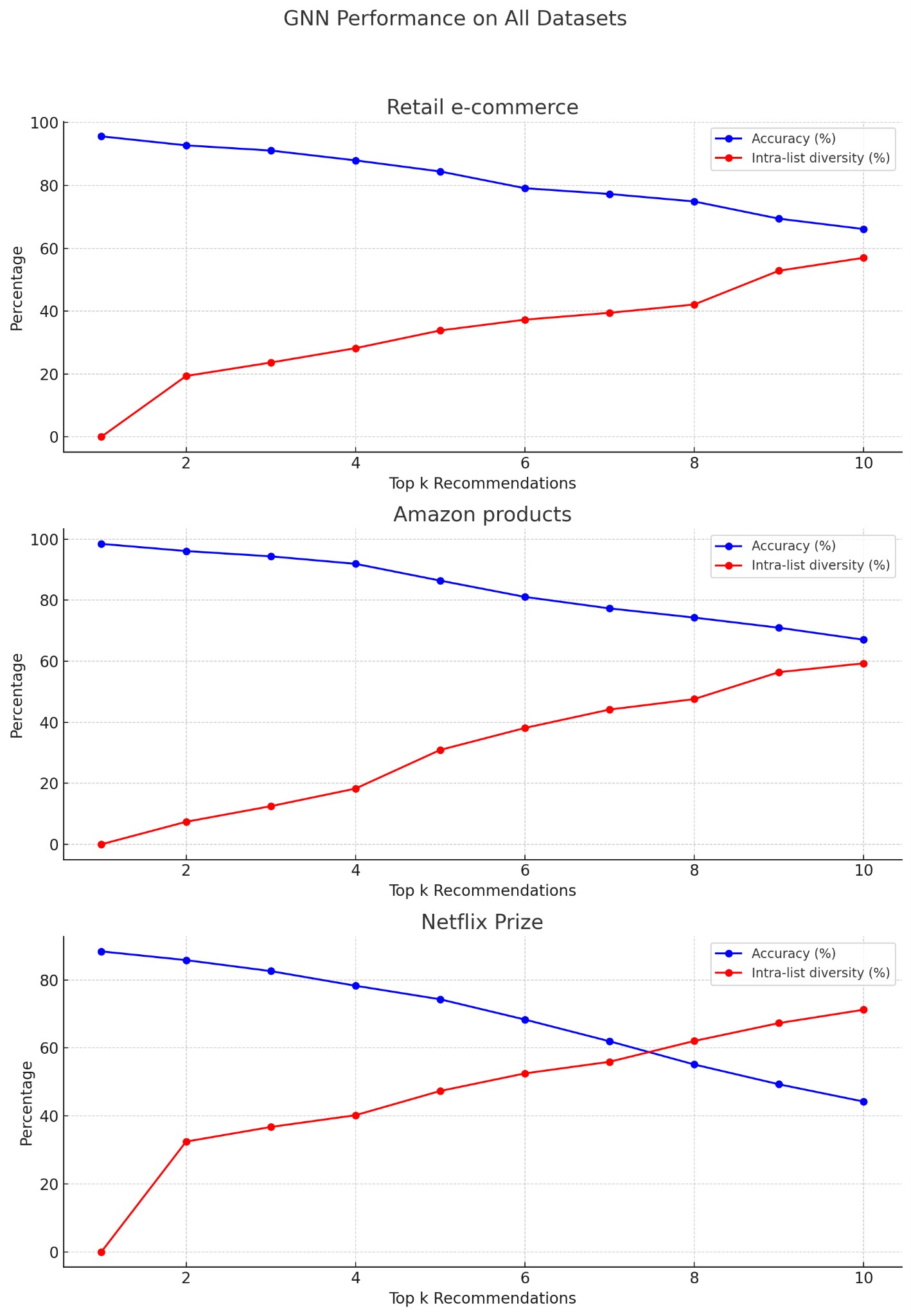}

\end{minipage}
\hfill
\begin{minipage}{0.49\textwidth}
    \centering
    \includegraphics[
  width=\linewidth,
  height=0.5\textheight,
  keepaspectratio
]{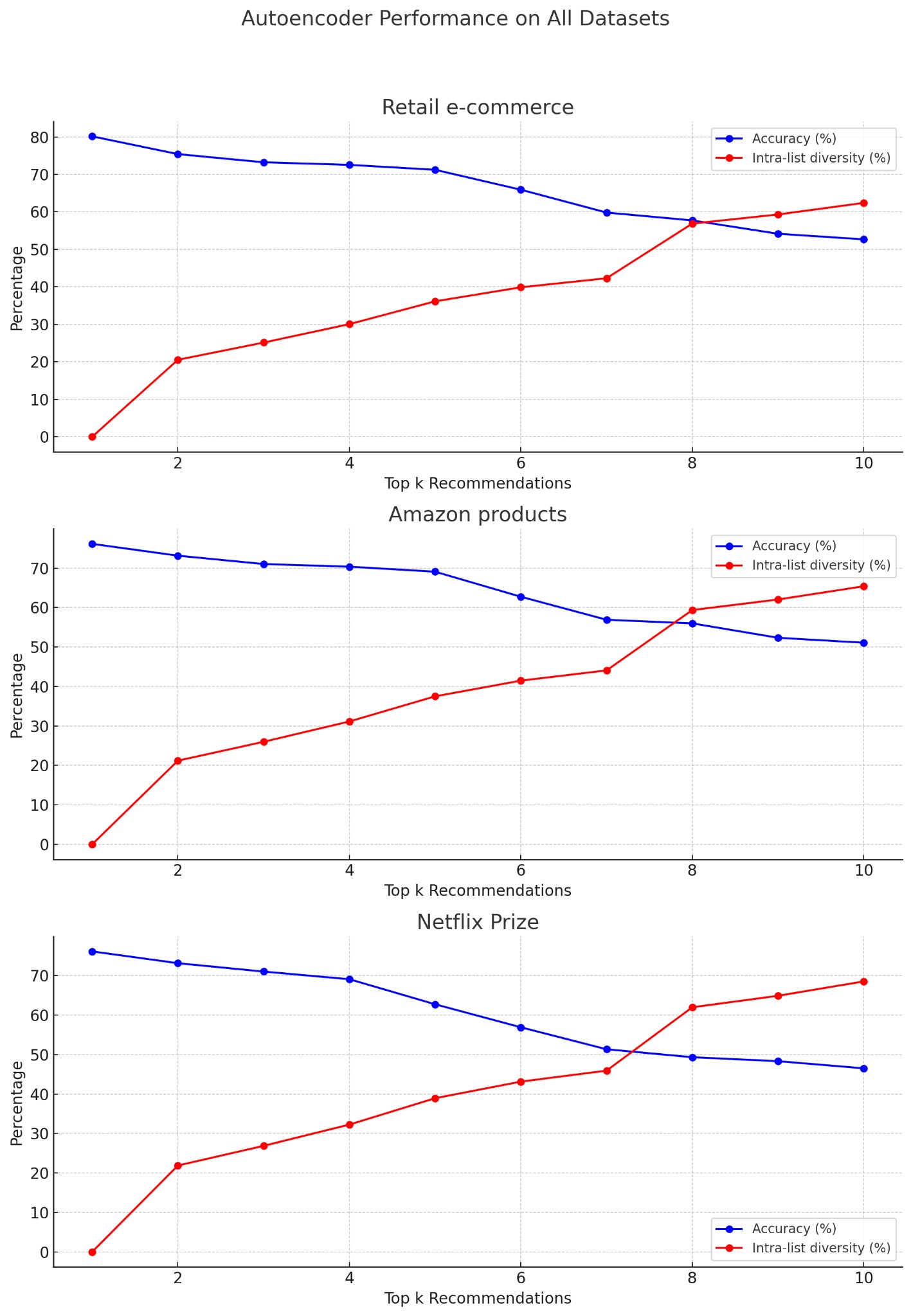}

\end{minipage}

\end{figure*}

\begin{figure*}[htbp!]
\centering

\begin{minipage}{0.49\textwidth}
    \centering
    \includegraphics[
  width=\linewidth,
  height=0.5\textheight,
  keepaspectratio
]{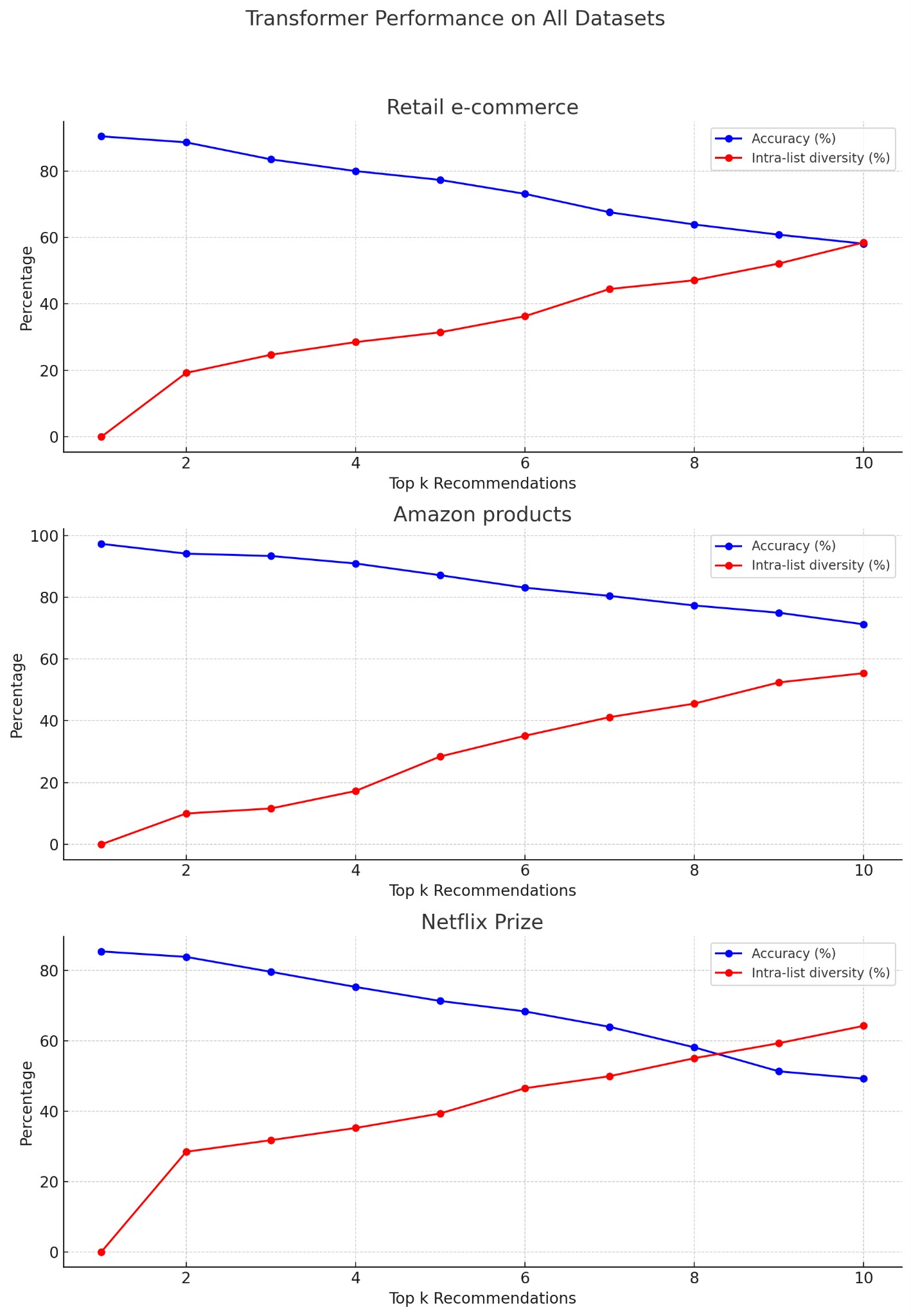}

\end{minipage}
\hfill
\begin{minipage}{0.49\textwidth}
    \centering
    \includegraphics[
  width=\linewidth,
  height=0.5\textheight,
  keepaspectratio
]{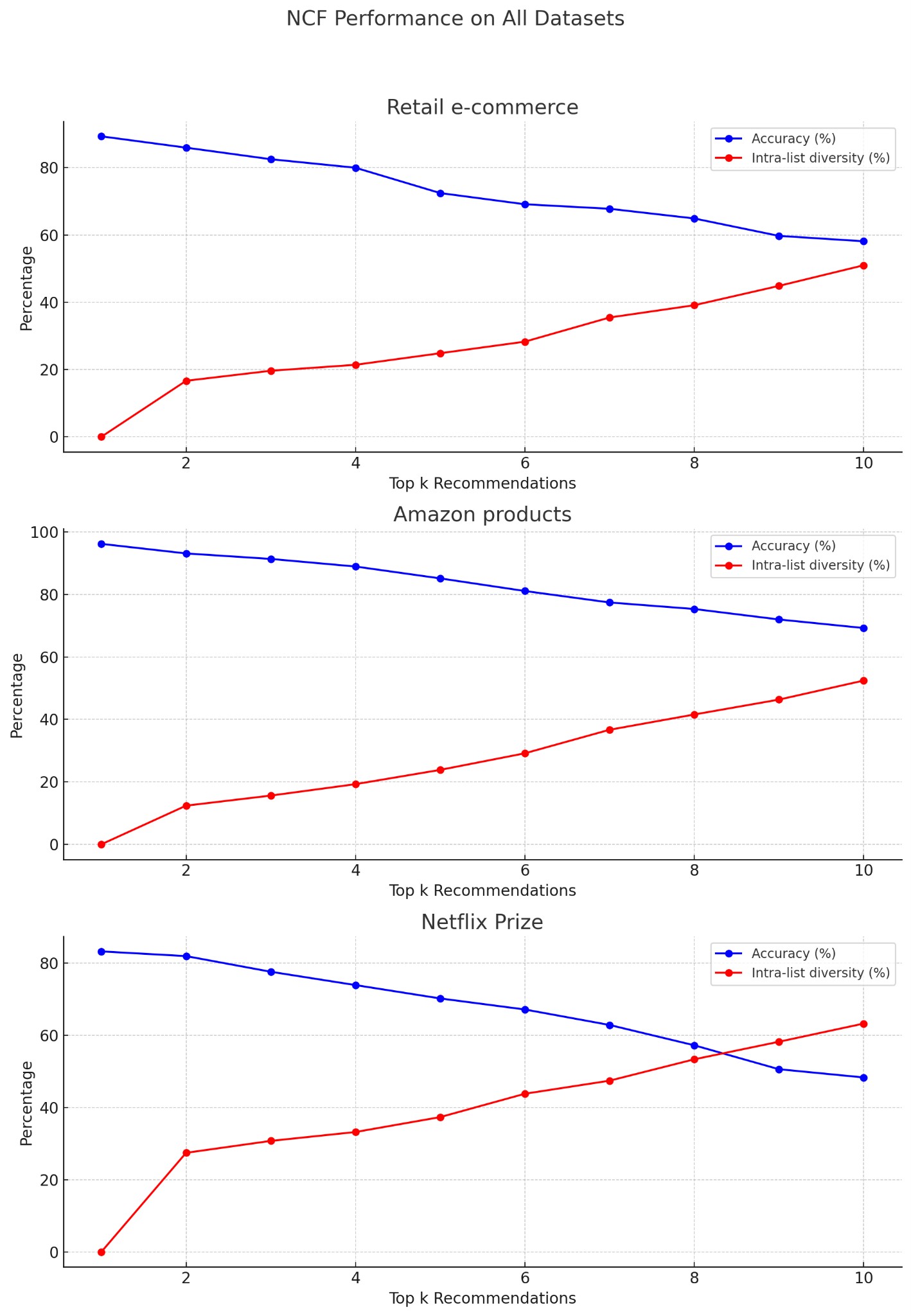}

\end{minipage}

\vspace{0.15cm}

\begin{minipage}{0.49\textwidth}
    \centering
    \includegraphics[
  width=\linewidth,
  height=0.5\textheight,
  keepaspectratio
]{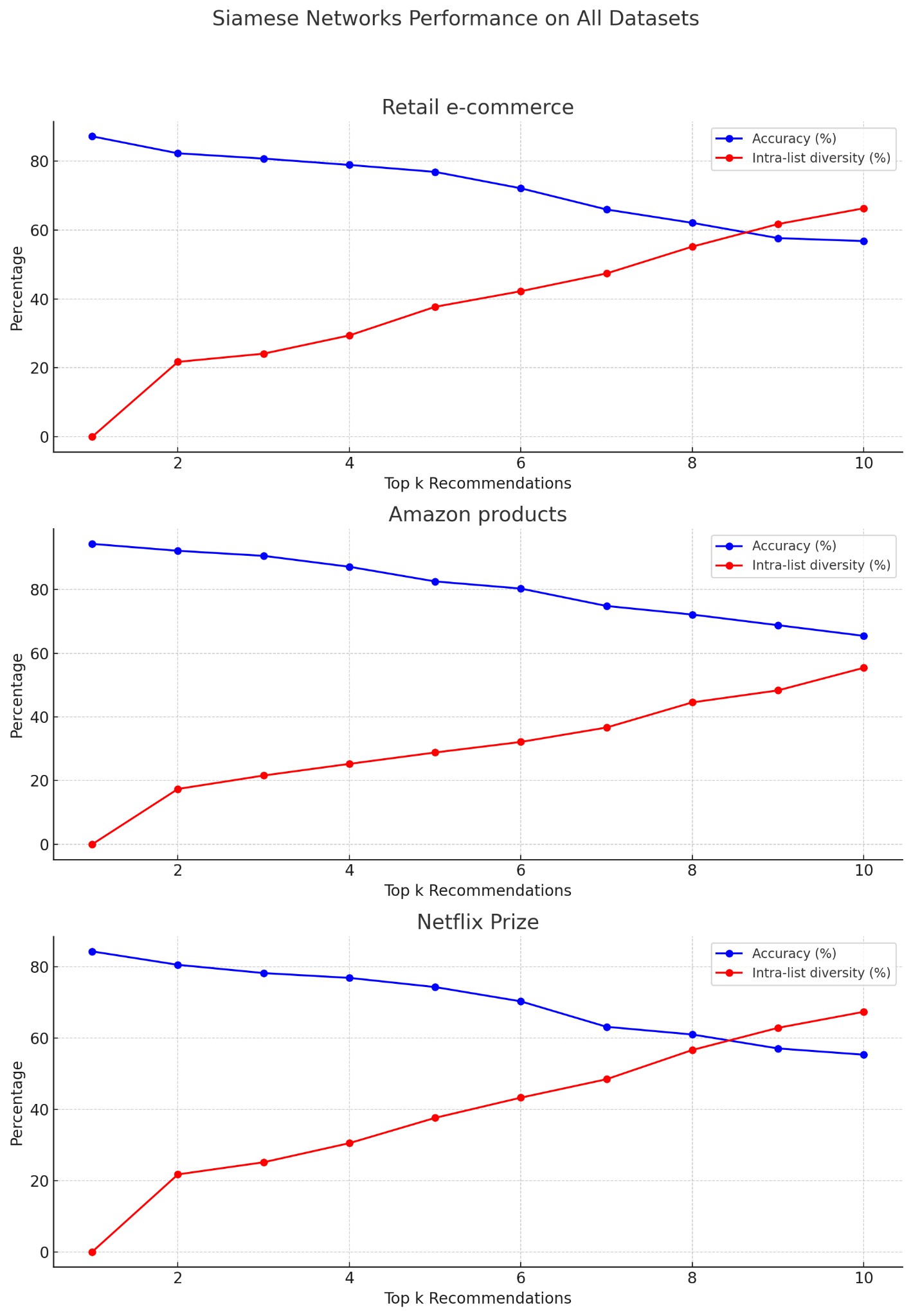}

\end{minipage}
\hfill

\caption{Performance comparison across datasets using multiple evaluation metrics.}
\label{fig:metrics_5_1_5_02}
\end{figure*}

\begin{figure*}[htbp!]
\centering

\begin{minipage}{0.49\textwidth}
    \centering
    \includegraphics[
  width=\linewidth,
  height=0.5\textheight,
  keepaspectratio
]{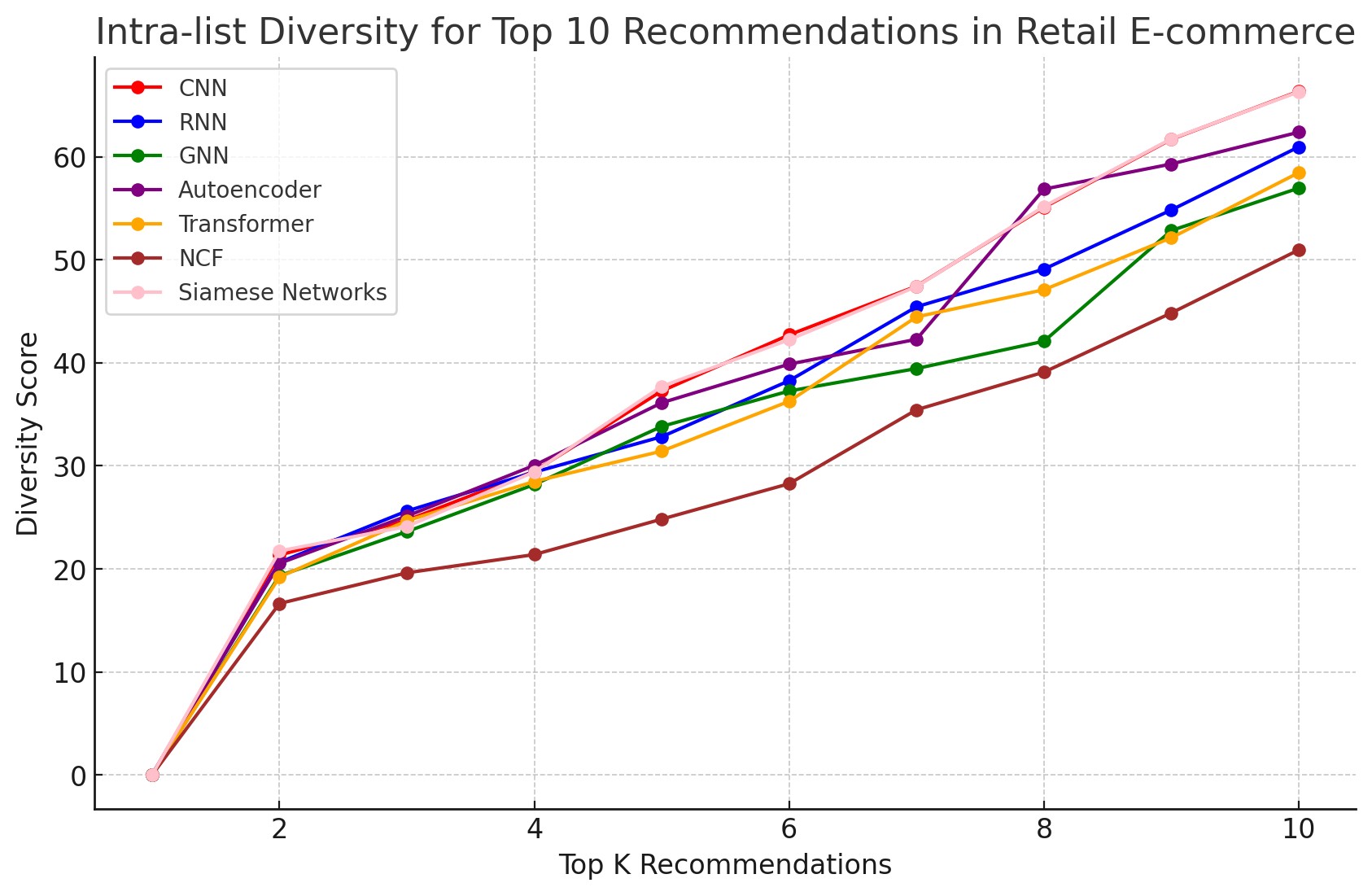}

\end{minipage}
\hfill
\begin{minipage}{0.49\textwidth}
    \centering
    \includegraphics[
  width=\linewidth,
  height=0.5\textheight,
  keepaspectratio
]{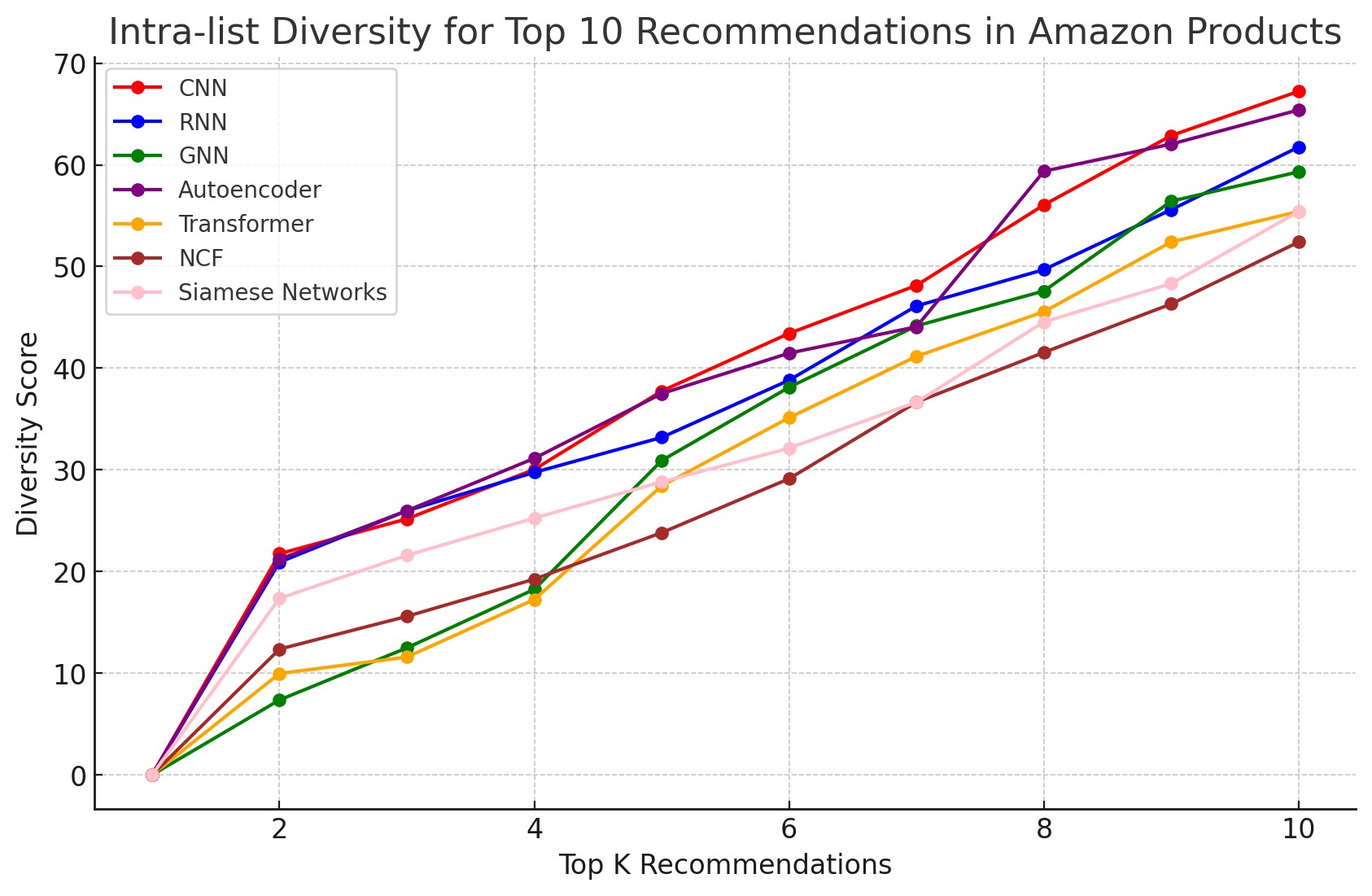}

\end{minipage}

\vspace{0.15cm}

\begin{minipage}{0.49\textwidth}
    \centering
    \includegraphics[
  width=\linewidth,
  height=0.5\textheight,
  keepaspectratio
]{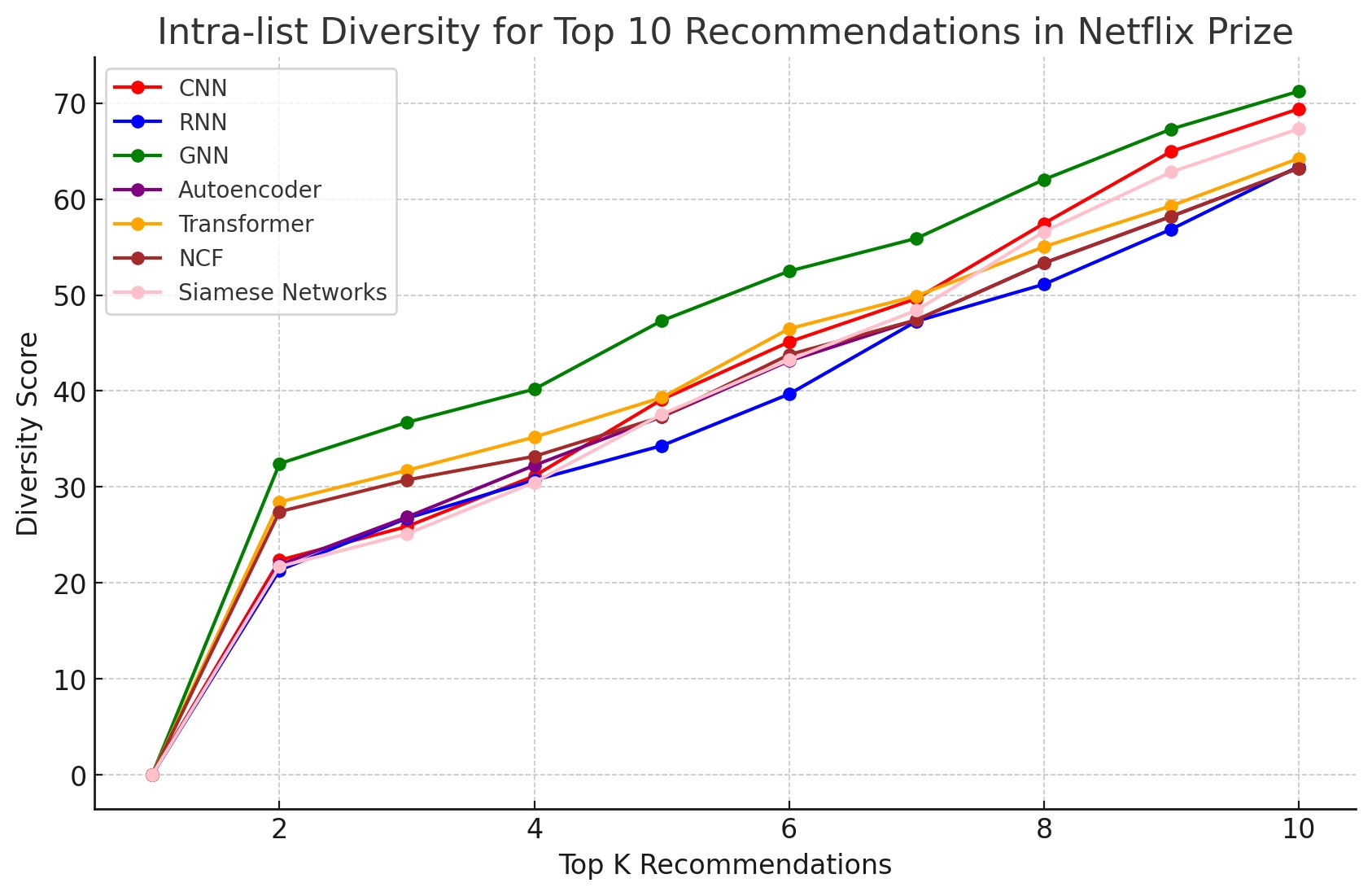}

\end{minipage}
\hfill

\caption{Performance comparison across datasets using multiple evaluation metrics.}
\label{fig:metrics_5_1_5_02}
\end{figure*}

\subsection{Computational Efficiency Analysis}
\begin{table*}[h!]
\centering
\small
\caption{Computational Cost Analysis of Recommendation Models Across Datasets}
\label{tab:computational_cost}
\resizebox{\textwidth}{!}{%
\begin{tabular}{l|ccc|ccc|ccc}
\hline
\multirow{2}{*}{\textbf{Model}} 
& \multicolumn{3}{c|}{\textbf{Retail Rocket}} 
& \multicolumn{3}{c|}{\textbf{Amazon Products}} 
& \multicolumn{3}{c}{\textbf{Netflix Prize}} \\
\cline{2-10}
& \textbf{Train (s)} & \textbf{Infer (ms)} & \textbf{Mem (MB)}
& \textbf{Train (s)} & \textbf{Infer (ms)} & \textbf{Mem (MB)}
& \textbf{Train (s)} & \textbf{Infer (ms)} & \textbf{Mem (MB)} \\
\hline
CNN 
& 420  & 6.2  & 380 
& 2\,850 & 9.5  & 820 
& 1\,120 & 7.1  & 510 \\

RNN 
& 680  & 9.8  & 420 
& 4\,200 & 14.3 & 960 
& 1\,950 & 11.6 & 620 \\

GNN 
& 910  & 12.4 & 540 
& 5\,800 & 18.7 & 1\,200 
& 2\,600 & 15.2 & 780 \\

Autoencoder 
& 360  & 5.1  & 310 
& 2\,100 & 7.8  & 650 
& 980  & 6.3  & 430 \\

Transformer 
& 1\,150 & 21.6 & 880 
& 7\,900 & 35.4 & 1\,650 
& 3\,400 & 27.9 & 1\,120 \\

NCF 
& 290  & 4.6  & 270 
& 1\,750 & 6.9  & 580 
& 820  & 5.4  & 390 \\

Siamese Networks 
& 520  & 8.2  & 360 
& 3\,100 & 12.1 & 740 
& 1\,480 & 9.7  & 560 \\
\hline
\end{tabular}%
}
\end{table*}
Table~\ref{tab:computational_cost} reports the training time, inference latency, and memory consumption of all evaluated architectures across the three datasets, addressing Requirement R5 (Computational Efficiency) of the ROB framework. The results reveal substantial variation in computational cost across model classes, reflecting differences in architectural complexity and scalability.

Lightweight architectures such as NCF and Autoencoders exhibit the lowest training times, inference latency, and memory usage across all datasets, indicating favorable efficiency characteristics for large-scale or resource-constrained deployments. CNN-based models demonstrate moderate computational cost, offering a balance between efficiency and representational capacity.
In contrast, GNNs and Transformer-based models incur significantly higher computational overhead, particularly on the Amazon Products and Netflix Prize datasets. This increase is attributed to graph message passing operations in GNNs and the quadratic complexity of self-attention mechanisms in Transformers. Recurrent Neural Networks and Siamese Networks occupy an intermediate position, with higher costs than convolutional and matrix-factorization-based approaches but lower overhead than attention- or graph-based architectures.

Overall, these findings highlight a clear trade-off between model expressiveness and computational efficiency. While complex architectures deliver strong predictive performance and richer relational modeling, their deployment may be constrained by training cost, inference latency, or memory requirements. This analysis underscores the importance of incorporating efficiency considerations alongside accuracy and diversity when selecting or combining models for real-world recommendation systems.

\subsection{Interpretation of Results}

The comparative evaluation of the seven neural architectures reveals that model performance is strongly influenced by dataset characteristics and underlying interaction patterns. Interpreted through the ROB framework, the results indicate that different architectures exhibit complementary strengths rather than universal dominance.

GNN achieve superior predictive accuracy on the Retail Rocket and Amazon datasets, reflecting their ability to exploit complex item–item relationships (R3) common in e- commerce environments. Transformer-based models also demonstrate competitive accuracy due to their capacity to capture global interaction patterns through attention mechanisms. In contrast, RNN outperform other architectures on the Netflix Prize dataset, highlighting their effectiveness in modeling sequential user behavior and temporal preference evolution.

With respect to recommendation diversity, Siamese Networks consistently yield higher intra-list diversity, particularly in retail-oriented datasets, owing to their similarity-based learning objective. CNN-based models also exhibit strong diversity characteristics in the Amazon dataset, while GNNs provide balanced diversity on Netflix due to rich relational connectivity. Overall, these findings emphasize that the choice of architecture should be guided by the desired balance between predictive accuracy and recommendation diversity, reinforcing the multi-objective nature of modern recommendation systems.
Based on the observed performance trends across datasets and evaluation requirements, Table~\ref{tab:rob_alignment} summarizes the empirical association between model classes and system requirements. It
\begin{table*}[h!]
\centering
\caption{Empirical alignment between neural network architectures and system requirements under the ROB framework. Avg TT: Average training time.}
\label{tab:rob_alignment}
\resizebox{\textwidth}{!}{%
\begin{tabular}{p{4cm} p{1.8cm} p{1.8cm} p{1.8cm} p{1.8cm} p{1.8cm}}
\toprule
\textbf{Model} & \textbf{R1} & \textbf{R2} & \textbf{R3} & \textbf{R4} & \textbf{R5} (Avg. TT) \\
\midrule
CNN & Moderate & High & Low & Moderate & Low \\
RNN & High & Low & High & Low & Moderate\\
GNN & High & Moderate & High & High & High \\
Autoencoder & Moderate & Low & Moderate & Low & Low \\
Transformer & High & Moderate & Low & High & High\\
NCF & High & Low & Low & Low & Low\\
Siamese Network & Moderate & Low & High & High & Low \\
\bottomrule
\end{tabular}
}
\end{table*}
 highlights that different neural architectures satisfy distinct system requirements under the ROB framework, with no single model excelling across all dimensions. This observation suggests that robust recommendation systems can benefit from ensemble or hybrid designs that integrate complementary modeling capabilities. For example, accuracy-oriented architectures such as GNNs, Transformers, or NCF can be combined with diversity-promoting models such as Siamese Networks or CNNs, while sequence-aware models like RNNs can enhance temporal adaptability. Such multi-model strategies provide a principled way to jointly address predictive accuracy, diversity, relational awareness, and temporal dynamics, offering a promising direction for building resilient and adaptable recommendation systems aligned with evolving application demands.

\subsection{Comparison with Related Work}
Existing research on recommendation systems has extensively explored deep learning approaches, including matrix factorization, conventional neural networks, and, more recently, graph-based and attention-based models. Prior studies typically evaluate these architectures in isolation or focus primarily on predictive accuracy.
In contrast, the present study adopts a requirement-oriented evaluation perspective that jointly considers accuracy, diversity, relational modeling capability, and temporal robustness. 
While GNN and Transformers have previously been shown to perform well in recommendation tasks, our findings further contextualize their strengths under a unified evaluation protocol across heterogeneous datasets. Similarly, the diversity-promoting behavior of Siamese Networks, which has been less emphasized in earlier benchmarking studies, is highlighted through explicit diversity analysis.

By systematically evaluating multiple architectures under consistent experimental conditions, this work complements existing literature by providing comparative insights that inform architecture selection and combination, rather than proposing alternative models or optimization strategies.

\subsection{Limitations}
Despite the promising results obtained under the ROB framework, several limitations should be acknowledged. Although this study reports quantitative measurements of training time, inference latency, and memory consumption (R5), the computational efficiency analysis is conducted under a fixed hardware and implementation setting. As a result, absolute efficiency values may vary across deployment environments, software optimizations, or hardware accelerators. Nevertheless, the reported results provide a consistent relative comparison of computational cost across architectures.
Model performance is also strongly influenced by data availability and quality. Sparse, noisy, or incomplete interaction data can degrade predictive accuracy and amplify bias, while large-scale data collection raises concerns related to privacy preservation and data security. Although sequence-aware architectures address temporal dynamics to some extent, the evaluated models are trained in offline settings and do not fully support real-time adaptation to abrupt shifts in user behavior or rapidly evolving trends, which remains a challenge in highly dynamic application domains.

In addition, achieving an effective balance between predictive accuracy (R1) and recommendation diversity (R2) remains non-trivial. Accuracy-oriented architectures may produce overly homogeneous recommendations, whereas diversity-focused approaches can reduce relevance. Addressing this trade-off requires further investigation into adaptive and hybrid strategies capable of dynamically adjusting system objectives based on contextual or user-driven signals. These limitations point toward future research directions that emphasize deployment-aware efficiency evaluation, robust data handling, and multi-objective optimization in real-world recommendation systems.

\section{Conclusion}

This paper presented a requirement-oriented benchmarking study of seven neural network architectures across three heterogeneous datasets, Retail E-commerce, Amazon Products, and Netflix Prize. By organizing evaluation around core system requirements, including predictive accuracy, recommendation diversity, relational awareness, temporal dynamics, and computational efficiency, the proposed ROB framework offers a unified and application-relevant perspective on modern recommendation system evaluation.
Experimental results show that no single architecture consistently satisfies all requirements. GNNs and Transformers achieve strong predictive accuracy in settings characterized by complex relational structures, while RNNs are particularly effective for modeling temporal user behavior. In contrast, Siamese Networks and CNN-based models contribute more prominently to recommendation diversity, emphasizing broader item exposure. These findings highlight the importance of aligning model selection with specific system objectives and data characteristics.
The analysis further reveals inherent trade-offs between model expressiveness and computational efficiency, as well as between predictive accuracy and recommendation diversity. Such trade-offs underscore the limitations of relying on a single architecture in real-world deployments and motivate the consideration of hybrid and ensemble strategies that integrate complementary modeling capabilities.
Overall, this work contributes a structured benchmarking perspective that complements existing recommendation system research by emphasizing multi-objective evaluation and practical design implications. The ROB framework provides a reusable reference for evaluating and designing recommendation systems that better align with the evolving demands of contemporary digital platforms. Looking ahead, we will explore adaptive ensemble mechanisms that dynamically balance accuracy, diversity, and efficiency under changing user and system constraints.
Future work may extend the ROB framework to evaluate large language model–based recommendation systems, where conversational reasoning, contextual understanding, and efficiency constraints introduce new evaluation challenges beyond traditional neural architectures.

\section*{Acknowledgment}
The authors would like to thank the editors and anonymous reviewers for their time and valuable comments.

\section*{Ethical Approval} 
Not Applicable.

\section*{Declaration of Vompeting interests}
The authors declare no conflict of interest.

\section*{Funding}
This research received no external funding.

\section*{Data Availability}
The data that support the findings of this study are available from the corresponding author upon reasonable request.


\end{document}